\documentclass[aps,prc,twocolumn,superscriptaddress,floatfix,amsmath,amssymb,nofootinbib,longbibliography]{revtex4-1}
\usepackage[utf8]{inputenc}
\usepackage[T1]{fontenc}

\usepackage{xcolor}
\usepackage{graphicx}
\usepackage[ocgcolorlinks,citecolor=blue,linkcolor=blue,urlcolor=blue]{hyperref}
\usepackage{scalefnt}
\usepackage{xspace}
\usepackage{times,txfonts}
\usepackage{pifont}
\usepackage{bm}

\newcommand{\la}{\langle}
\newcommand{\ra}{\rangle}

\newcommand{\hw}{$\hbar \omega$\xspace}
\newcommand{\Msmall}{\mathcal{M}_\text{reduced}}
\newcommand{\Mlarge}{\mathcal{M}_\text{full}}
\newcommand{\ai}{\emph{ab initio}}
\newcommand{\tr}{\ensuremath{\text{tr}}}
\newcommand{\ame}[2]{\ensuremath{  \Gamma_{#1#2} }}
\newcommand{\pr}{\prime}
\newcommand{\HF}{\text{HF}}
\newcommand{\NAT}{\text{NAT}}
\newcommand{\op}[3]{\ensuremath #1^{(#2 \text{B})}_{#3}}
\newcommand{\opno}[3]{\ensuremath \tilde #1^{(#2 \text{B})}_{#3}}

\begin{document}

\title{Natural orbitals for many-body expansion methods}

\author{J.~Hoppe}
\email{jhoppe@theorie.ikp.physik.tu-darmstadt.de}
\affiliation{Technische Universit\"at Darmstadt, Department of Physics, 64289 Darmstadt, Germany}
\affiliation{ExtreMe Matter Institute EMMI, GSI Helmholtzzentrum f\"ur Schwerionenforschung GmbH, 64291 Darmstadt, Germany}

\author{A.~Tichai}
\email{alexander.tichai@physik.tu-darmstadt.de}
\affiliation{Max-Planck-Institut f\"ur Kernphysik, Saupfercheckweg 1, 69117 Heidelberg, Germany}
\affiliation{Technische Universit\"at Darmstadt, Department of Physics, 64289 Darmstadt, Germany}
\affiliation{ExtreMe Matter Institute EMMI, GSI Helmholtzzentrum f\"ur Schwerionenforschung GmbH, 64291 Darmstadt, Germany}

\author{M.~Heinz}
\email{mheinz@theorie.ikp.physik.tu-darmstadt.de}
\affiliation{Technische Universit\"at Darmstadt, Department of Physics, 64289 Darmstadt, Germany}
\affiliation{ExtreMe Matter Institute EMMI, GSI Helmholtzzentrum f\"ur Schwerionenforschung GmbH, 64291 Darmstadt, Germany}

\author{K.~Hebeler}
\email{kai.hebeler@physik.tu-darmstadt.de}
\affiliation{Technische Universit\"at Darmstadt, Department of Physics, 64289 Darmstadt, Germany}
\affiliation{ExtreMe Matter Institute EMMI, GSI Helmholtzzentrum f\"ur Schwerionenforschung GmbH, 64291 Darmstadt, Germany}

\author{A.~Schwenk}
\email{schwenk@physik.tu-darmstadt.de}
\affiliation{Technische Universit\"at Darmstadt, Department of Physics, 64289 Darmstadt, Germany}
\affiliation{ExtreMe Matter Institute EMMI, GSI Helmholtzzentrum f\"ur Schwerionenforschung GmbH, 64291 Darmstadt, Germany}
\affiliation{Max-Planck-Institut f\"ur Kernphysik, Saupfercheckweg 1, 69117 Heidelberg, Germany}

\begin{abstract}
The nuclear many-body problem for medium-mass systems is commonly addressed using wave-function expansion methods that build upon a second-quantized representation of many-body operators with respect to a chosen computational basis.
While various options for the computational basis are available, perturbatively constructed natural orbitals recently have been shown to lead to significant improvement in many-body applications yielding faster model-space convergence and lower sensitivity to basis set parameters in large-scale no-core shell model diagonalizations.
This work provides a detailed comparison of single-particle basis sets and a systematic benchmark of natural orbitals in nonperturbative many-body calculations using the in-medium similarity renormalization group approach.
As a key outcome we find that the construction of natural orbitals in a large single-particle basis enables for performing the many-body calculation in a reduced space of much lower dimension, thus offering significant computational savings in practice that help extend the reach of \ai{} methods towards heavier masses and higher accuracy.
\end{abstract}



\maketitle

\section{Introduction}
\label{sec:intro}

Nuclear many-body theory has witnessed major developments over the last two decades, extending the reach of the \ai{} solution of the stationary Schr\"odinger equation over a wide range of mass numbers in the nuclear chart, covering closed- and open-shell nuclei and including exotic nuclei~\cite{Hebe15ARNPS,Herg20review}.
This progress is mainly based on (i) the construction of improved nucleon-nucleon (\textit{NN}) and three-nucleon (3\textit{N}) interactions based on chiral effective field theory (EFT)~\cite{Epel09RMP,Mach11PR,Hebe11fits,Ekst15sat,Epel15NNn4lo,Ente17EMn4lo,Dris17MCshort,Hopp19medmass,Huth19chiralfam,Epel19nuclfFront,Hebe203NF,Jian20N2LOGO} and (ii) the extension of many-body theories applicable to medium-mass nuclei~\cite{Dick04PPNP,Hage14RPP,Herg16PR,Tichai2020review}.
The advances of many-body calculations are intimately linked to the use of wave-function expansion methods, which exhibit mild computational scaling in mass number, instead of the exponential scaling required by exact methods, resulting in the recent milestone \ai{} calculation of $^{100}$Sn~\cite{Morr17Tin}.
In practice, various many-body approaches exist for medium-mass nuclei, e.g., many-body perturbation theory (MBPT)~\cite{Holt14Ca,Tich16HFMBPT,Tichai18BMBPT,Tichai2020review}, coupled cluster (CC) theory~\cite{Hage14RPP,Bind14CCheavy}, the in-medium similarity renormalization group (IMSRG)~\cite{Tsuk11IMSRG,Herg16PR,Stroberg2019}, self-consistent Green's function (SCGF) theory~\cite{Dick04PPNP,Soma20SCGF}, and nuclear lattice simulations~\cite{Lahd13LEFT}. 
In particular, the recent use of nonperturbative many-body approaches has generated an unprecedented level of accuracy in medium-mass applications for a diverse set of nuclear observables (see, e.g., Refs.~\cite{Hage16NatPhys,Gysb19beta,Yao2020}).

All these frameworks require the introduction of a computational basis for the representation of the (second-quantized) many-body operators.
In the limit of a one-body Hilbert space of infinite dimension, different choices of the computational basis yield identical results.
However, due to computational limitations, in practice one is always restricted to using a finite basis size and, consequently, the resulting observables will depend on the underlying computational basis.

It has been realized only recently in nuclear physics that the optimization of the single-particle basis provides a powerful tool to stabilize many-body calculations and enables a more reliable extraction of physical observables from large-scale calculations~\cite{Capr12Sturmian,Tich19NatNCSM}.
In other fields of many-body research, like quantum chemistry, this is much more explored and the construction of suitable single-particle basis functions has been an integral part of the \ai{} endeavor, yielding a rich variety of basis sets.
In this work, we investigate benefits and limitations of various single-particle bases used in the solution of the nuclear many-body problem.

Choosing the single-particle basis in nuclear many-body theory primarily requires addressing the following questions:
\begin{itemize}
\item[(i)] What is the best choice for obtaining rapid convergence with respect to the model-space size?
\item[(ii)] What is the best strategy to minimize the dependence of physical observables on basis set parameters?
\item[(iii)] To what extent is the factorization of center-of-mass and intrinsic motion contaminated?
\end{itemize}
In practice, optimizing with respect to all of the above simultaneously is not possible.
Historically, most many-body calculations either employ harmonic oscillator (HO) or Hartree-Fock (HF) single-particle states.
Harmonic oscillator basis states rigorously ensure factorization of center-of-mass and intrinsic degrees of freedom of the many-body wave function when combined with an $N_\text{max}$ truncation, as in no-core shell model (NCSM) approaches~\cite{Navr09NCSMdev,Barr13PPNP}. 
However, in practice a strong dependence on the basis set parameters such as the oscillator frequency of the confining potential is observed, especially for heavier nuclei or for observables that are more sensitive to the long-range part of the nuclear wave function.
This makes the extraction of such observables challenging.
Using HF orbitals based on a prior mean-field solution typically lowers the frequency dependence, 
while numerically still leading to a factorization of the center-of-mass and intrinsic wave function in large enough model spaces~\cite{Hage09CoM}.
However, selected nuclear observables may still show sensitivity to the oscillator frequency in the HF basis as observed, e.g., for charge radii of medium-mass nuclei in IMSRG calculations~\cite{Hopp19medmass}.

Recently, applications of natural orbitals (NAT), defined as eigenvectors of the one-body density matrix, revealed faster model-space convergence and significantly reduced sensitivity to basis parameters in large-scale NCSM calculations~\cite{Tich19NatNCSM}.
Furthermore, they have been shown to drastically reduce the required amount of three-particle-three-hole amplitudes in CC applications~\cite{Novario2020a}, allowing for novel calculations with leading triples corrections, e.g., for deformed nuclei~\cite{Novario2020a} and nuclear matrix elements of the neutrinoless double-$\beta$ decay~\cite{Novario2020b}.
In this work, the natural orbital basis is inspected in detail and systematically applied in nonperturbative medium-mass studies using modern chiral interactions.

This paper is organized as follows.
In Sec.~\ref{sec:basisopt} various options of single-particle bases are shown and discussed.
Section~\ref{sec:normalordering} introduces the concept of normal ordering as well as the impact of the reference state on the many-body formalism.
In Sec.~\ref{sec:imsrg} the IMSRG approach is briefly introduced.
The different single-particle bases considered are compared in detail in Sec.~\ref{sec:results1} and applied to medium-mass systems using the IMSRG formalism in Sec.~\ref{sec:results2}.
Finally, we summarize and conclude in Sec.~\ref{sec:outlook}.

\section{Basis optimization}
\label{sec:basisopt}

\subsection{Rationale}

While HO basis sets have been used extensively for a long time in various many-body frameworks, they constitute an agnostic choice with respect to any specific properties of the target system, e.g., in terms of mass number or mean-field effects.
This can be addressed by using HF orbitals instead.
Hartree-Fock orbitals account for bulk properties of the nucleus stemming from a variational minimization of the ground-state energy.
Observables like the energy or the radius are therefore well captured at the HF level as long as the nuclear interaction is soft enough, and the single-particle wave functions possess an improved radial dependence as opposed to the Gaussian falloff of HO eigenfunctions.
Proton and neutron single-particle potentials in general differ in the HF approach, thus accounting for mean-field contributions induced by Coulomb and isospin-breaking effects.

Still, the HF procedure by construction only provides an optimization of occupied single-particle states (holes) while leaving virtual single-particle states (particles) untouched beyond fixing the normalization.\footnote{Unitary transformations applied separately to the hole and particle space leave the HF energy invariant, but impact the single-particle energies and wave functions (see also Sec.~\ref{sec:tkin}).}
However, wave-function expansion methods aim at capturing dynamic correlations linked to particle-hole excitations, which also involve single-particle orbitals that are not optimized by the HF approach.
Therefore, incorporating such effects in the construction of the computational basis is key when trying to robustly determine observables to high precision.

\subsection{Notation}

In the following, we denote second-quantized $n$-body ($n$B) operators via
\begin{align}
    \op{O}{n}{} \equiv \frac{1}{(n!)^2} \sum_{k_1 \cdots k_{2n}} \op{o}{n}{k_1 \cdots k_{2n}} 
    a^\dagger_{k_1} \cdots a^\dagger_{k_n} a_{k_{2n}} \cdots a_{k_{n+1}}\,,
\end{align}
where the lower-case letters $\op{o}{n}{k_1 \cdots k_{2n}}$ represent their matrix elements and $a^\dagger_k$ ($a_k$) denote the single-particle creation (annihilation) operators.
Normal-ordering techniques are exploited to reexpress the operator with respect to an $A$-body reference state,
\begin{align}
    \opno{O}{n}{} \equiv \frac{1}{(n!)^2} \sum_{k_1 \cdots k_{2n}} \opno{o}{n}{k_1 \cdots k_{2n}} 
   :a^\dagger_{k_1} \cdots a^\dagger_{k_n} a_{k_{2n}} \cdots a_{k_{n+1}}:\,,
\end{align}
where strings of normal-ordered creation and annihilation operators are denoted by colons and we use the tilde to distinguish the reference-state normal-ordered operator and its matrix elements from the initial one.
While the specific vacuum is absent in this notation, it will be clear from the context what reference state we are referring to.

For the particular case of the nuclear Hamiltonian the following notation is employed to denote its normal-ordered contributions:
\begin{align}
    H &= E_0 + \sum_{pq} f_{pq} :a^\dagger_p a_q: \notag \\
    &\phantom{=}+ \frac{1}{4} \sum_{pqrs} \Gamma_{pqrs} :a^\dagger_p a^\dagger_q a_s a_r: \notag \\
    &\phantom{=}+ \frac{1}{36} \sum_{pqrstu} W_{pqrstu} :a^\dagger_p a^\dagger_q a^\dagger_r a_u a_t a_s: \,, 
    \label{eq:hamno}
\end{align}
where $E_0$, $f$, $\Gamma$, and $W$ denote the zero-, one-, two-, and three-body matrix elements.
Because the operator in Eq.~\eqref{eq:hamno} is in reference-state normal order the expectation value is given by its normal-ordered zero-body part
\begin{align}
    \la \Phi | H | \Phi \ra = E_0 \,,
\end{align}
where $| \Phi \ra$ denotes the reference state.
In the case of a Hartree-Fock reference state $|\Phi\ra = |\HF\ra$ this corresponds to the Hartree-Fock mean-field energy, $E_0 = E_{\text{HF}}$.

\subsection{Natural orbitals}

Natural orbitals are defined as the eigenbasis of the one-body density matrix with its matrix elements given by
\begin{align}
\gamma_{pq} \equiv \frac{ \la \Psi | : a^\dagger_p a_q : | \Psi \ra}{\la \Psi | \Psi \ra} \, ,
\end{align}
where $|\Psi \ra$ denotes the exact ground state.
The calculation of the exact one-body density matrix requires the full solution of the Schr\"odinger equation, which is out of reach beyond the lightest systems.
However, early attempts in quantum chemistry revealed that \emph{approximate natural orbitals} can be very useful~\cite{Hay73natPT,Siu74CInatPT}.
Such basis sets are obtained by using an approximate many-body state $|\Psi^\text{approx} \ra$ to obtain an approximate one-body density matrix,
\begin{align}
\gamma^\text{approx}_{pq} \equiv \frac{ \la \Psi^\text{approx} | \, : a^\dagger_p a_q : \, | \Psi^\text{approx} \ra}{\la \Psi ^\text{approx}| \Psi^\text{approx.} \ra} \, .
\end{align}
In regions of the nuclear chart where the exact wave function is computationally inaccessible, this provides an alternative option for defining a basis for the many-body calculation. In practice, a reasonable trade-off between the accuracy of the many-body truncation for the construction of the approximate wave function and the associated computational cost needs to be found.

In the case where the approximate wave function is an HF Slater determinant $| \HF \ra$ the density matrix
\begin{align}
\gamma^{\text{HF}}_{pq} &\equiv \frac{ \la \HF | : a^\dagger_p a_q : | \HF \ra}{\la \HF | \HF \ra} \, 
\end{align}
has the particularly simple form
\begin{align}
\gamma^{\text{HF}}    
=
\begin{pmatrix}
\bm{1}_{\text{hh}} & 0 \\
0 & 0
\end{pmatrix} \, ,
\label{eq:rho_HF}
\end{align}
where $\bm{1}_{\text{hh}}$ denotes the identity in the sub-block of hole states.
As the HF state is normalized to unity,
the presence of the mean-field overlap $\la \HF | \HF \ra $ does not affect the HF density matrix.
Furthermore, the density matrix corresponds to a normalized many-body state,
meaning that its trace yields the particle number of the state, i.e., $\text{tr}(\gamma^{\text{HF}}) = A$.

For an HF reference state, the \emph{canonical orbitals}, defined as the eigenbasis of the one-body HF Hamiltonian, and the natural orbitals based on the HF density matrix in Eq.~\eqref{eq:rho_HF} coincide.
Therefore, one must include correlations beyond mean field in the construction of the density matrix to gain a benefit from the natural orbitals.

\subsection{Perturbatively improved density matrix}

\begin{figure}[t!]
\centering
\includegraphics[width=\columnwidth,clip=]{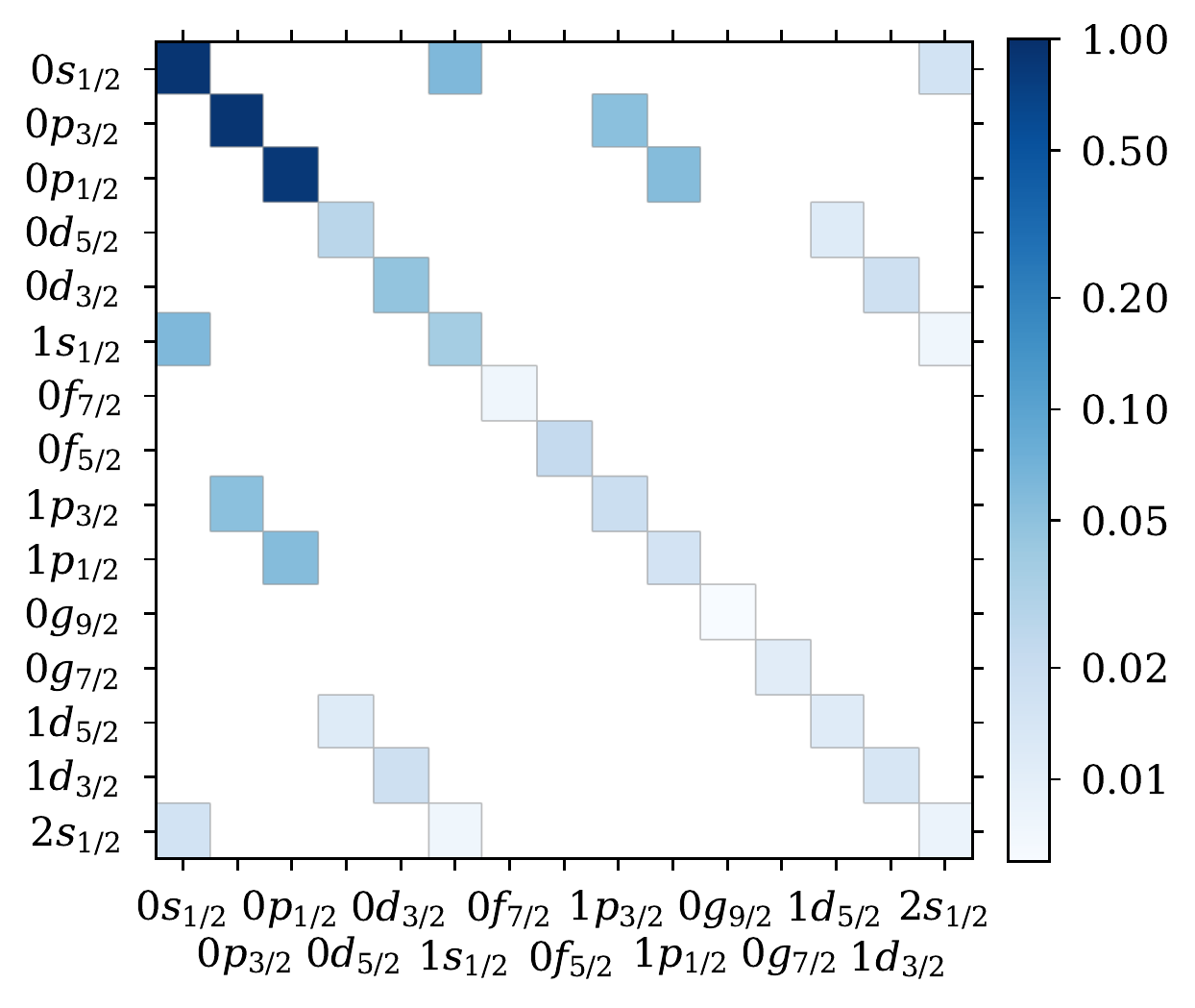}
\caption{
\label{fig:rho_MP2_contour}
Correlated one-body proton density matrix $\gamma^\text{MP2}$ in the HF basis for $^{16}$O in an $e_\text{max}=4$ model space using the next-to-next-to-next-to-leading order (N$^3$LO 450) interaction
based on the N$^3$LO \textit{NN} potential from Ref.~\cite{Ente17EMn4lo} with N$^3$LO 3\textit{N} forces constructed in Ref.~\cite{Dris17MCshort}.
The first three proton orbitals are occupied in the $^{16}$O reference state, while the remaining ones are unoccupied.
The perturbative corrections beyond HF can be seen in the diagonal particle-particle contributions and the off-diagonal particle-hole and hole-particle contributions. Note that for this $N=Z$ nucleus the neutron density matrix is very similar.}
\end{figure}

As discussed in the previous section, accounting for particle-hole couplings in the density matrix is essential for providing a more refined computational basis.
The simplest approach to including such effects is by employing a perturbatively corrected one-body density matrix.
Following the description in Ref.~\cite{Stra73nucldens}, the one-body density matrix up to second order in the interaction ($\lambda^2$), based on expanding the eigenstate of the approximate wave function up to second order in MBPT (MP2), can be written as
\begin{align}
\label{eq:diags}
\gamma^{\text{MP2}} \equiv \gamma^{\text{HF}} + \gamma^{(02)} + \gamma^{(20)} + \gamma^{(11)} + \mathcal{O}(\lambda^3)\, ,
\end{align}
where
\begin{align}
\gamma^{(mn)}_{pq} \equiv  \la \Phi^{(m)} | :a^\dagger_p a_q : |\Phi^{(n)} \ra \,  
\end{align}
is the MBPT contribution for the density matrix arising from the bra and ket wave function corrections at orders $m$ and $n$, respectively.
Terms of order $\lambda^3$ or higher in the interaction are discarded.
Moreover, terms of the form $\gamma^{(01)/(10)}$ are absent when using a canonical HF reference state due to Brillouin's theorem~\cite{Szab89QChem}.
Explicit expressions for the various contributions in terms of single-particle orbitals are given by
\begin{subequations}
\begin{align}
\label{eq:DAhp}
D^{\text{(hp)}_1}_{i^\pr a^\pr} &= \phantom{-}\frac{1}{2}\sum_{abi}  \frac{\ame{i^\pr i}{ab} \ame{ab}{a^\pr i} } { \epsilon_{i^\pr}^{a^\pr}  \epsilon_{i^\pr i}^{ab} } \, , \\
\label{eq:DBhp}
D^{\text{(hp)}_2}_{i^\pr a^\pr} &= -\frac{1}{2}\sum_{aij} \frac{\ame{i^\pr a}{ij} \ame{ij}{a^\pr a} } { \epsilon_{i^\pr}^{a^\pr} \epsilon_{i j}^{a^\pr a} } \, , \\
\label{eq:DChh}
D^\text{(hh)}_{i^\pr j^\pr} &= - \frac{1}{2} \sum_{abi} \frac{\ame{i^\pr i}{ab} \ame{ab}{j^\pr i} } { \epsilon_{i^\pr i}^{a b} \epsilon_{j^\pr i}^{ab} }\, , \\
\label{eq:DDpp}
D^\text{(pp)}_{a^\pr b^\pr } &= \phantom{-} \frac{1}{2} \sum_{aij} \frac{\ame{a^\pr a}{ij} \ame{ij}{b^\pr a} } { \epsilon_{ij}^{a^\pr a} \epsilon_{ij}^{b^\pr a} } \, ,
\end{align}%
\label{eq:mbptdens}%
\end{subequations}
where the labels $i,j,k,...$ ($a,b,c,...)$ correspond to single-particle states occupied (unoccupied) in the reference determinant, i.e., the HF state in our case.
The matrix elements $\ame{pq}{rs}$ are given in the HF basis, thus corresponding to an HF partitioning in the MBPT expansion of the density matrix~\cite{Tich16HFMBPT}.
Furthermore, the shorthand notation
\begin{align}
\epsilon^{ab}_{ij} \equiv \epsilon_{i} + \epsilon_{j} - \epsilon_{a} - \epsilon_{b} \,
\end{align}
is used, with $\epsilon_{p} \equiv f_{pp}$ denoting the HF single-particle energy of orbital $p$.
Consequently, the MP2 density matrix is given by
\begin{align}
\gamma^{\text{MP2}} =
\begin{pmatrix}
\gamma^{\text{hh}} & \gamma^{\text{hp}} \\
\gamma^{\text{ph}} & \gamma^{\text{pp}}
\end{pmatrix} \, ,
\label{eq:gamma_MP2}
\end{align}
where the hole-particle and particle-hole blocks are nonzero and given by
\begin{align}
\gamma^{\text{hp}} = D^{\text{(hp)}_1} + D^{\text{(hp)}_2} = \left(\gamma^{\text{ph}}\right)^\intercal \,,
\end{align}
and the hole-hole and particle-particle blocks by
\begin{subequations}
\begin{align}
\gamma^{\text{hh}} &= \gamma^\text{HF} + D^\text{(hh)} \,, \\
\gamma^{\text{pp}} &= D^\text{(pp)} \,,
\end{align}%
\end{subequations}
respectively.
Note that in contrast to the HF density matrix, the MP2 density matrix contains particle-particle and particle-hole couplings as shown for $^{16}$O in Fig.~\ref{fig:rho_MP2_contour}.
Since
\begin{align}
\sum_i D_{ii}^\text{(hh)} + \sum_a D_{aa}^\text{(pp)} =0\, ,
\end{align}
the second-order density matrix still fulfills the trace normalization condition $\tr(\gamma^\text{MP2})=A$ as in the HF case.

In practice, the construction of the MP2 density matrix is realized using a spherically constrained scheme, i.e., enforcing angular-momentum conservation throughout the initial HF solution and the following MBPT calculation.
Specifically, the single-particle orbitals are then characterized by the quantum numbers $n$, $l$, $j$, and $t$,
which are $(2j+1)$-fold degenerate. Here $n$ is the radial quantum number, $l$ the orbital angular momentum, $j$
the total angular momentum, and $t$ the isospin projection.
In actual calculations, we truncate the single-particle states at $e \leq e_\text{max}$, with quantum numbers $e=2n+l$.

Consequently, the resulting MP2 density matrix is block diagonal in the quantum numbers $ljt$ as only states with different radial quantum number $n$ couple.
The diagonalization of the MP2 density matrix is performed in sub-blocks to ensure symmetry conservation.
The resulting eigenvectors and eigenvalues correspond to the transformation coefficients from the HF to the NAT basis and the occupation numbers of the natural orbitals, respectively.

\subsection{Basis transformation}
\label{sec:basis_trafo}

The natural orbital states are obtained as linear combinations of the HF states, mixing radial excitations only:
\begin{align}
| n \alpha_p \ra_\text{NAT} = \sum_{n^\pr} \ ^{\text{NAT}}C_{n n^\pr}^{\alpha_p} \
| n^\pr \alpha_p \ra_\text{HF} \,,
\label{eq:CNat}
\end{align}
where $\alpha_p$ is a collective index for the quantum numbers $l_p$,~$j_p$, and $t_p$ and 
$^\text{NAT}C_{n n^\pr}^{\alpha_p} $ denotes the expansion coefficients in the HF basis obtained by the diagonalization, i.e.,
\begin{align}
_\text{HF} \la  n^\pr \alpha_p  | n \alpha_p \ra_\text{NAT} &=  \,^{\text{NAT}}C_{n n^\pr}^{\alpha_p} \,,
\end{align}
where the $m$ projection of the total angular momentum is suppressed since the transformation coefficients and single-particle states do not depend on it as long as rotational symmetry is enforced.
By expanding the HF states in the HO basis, we can also express the natural orbital states in the HO basis:
\begin{align}
| n \alpha_p \ra_\text{NAT} &= \sum_{n^\pr n^{\pr \pr}}  \ ^{\text{NAT}}C_{n n^\pr}^{\alpha_p}
\ ^{\text{HF}} C_{n^\pr n^{\pr \pr}}^{\alpha_p}
| n^{\pr \pr} \alpha_p \ra_\text{HO} \notag \\
&= \sum_{n^{\pr \pr}} \ ^{\text{NAT/HF}}C_{n n^{\pr \pr}}^{\alpha_p}
| n^{\pr \pr} \alpha_p  \ra_\text{HO} \,,
\label{eq:CNatHF}
\end{align}
where the coefficients $^{\text{NAT/HF}}C_{n n^{\pr \pr}}^{\alpha_p}$ now combine the transformation from the HO to the HF and from the HF to the NAT basis.

\begin{figure}[t!]
\centering
\includegraphics[width=\columnwidth,clip=]{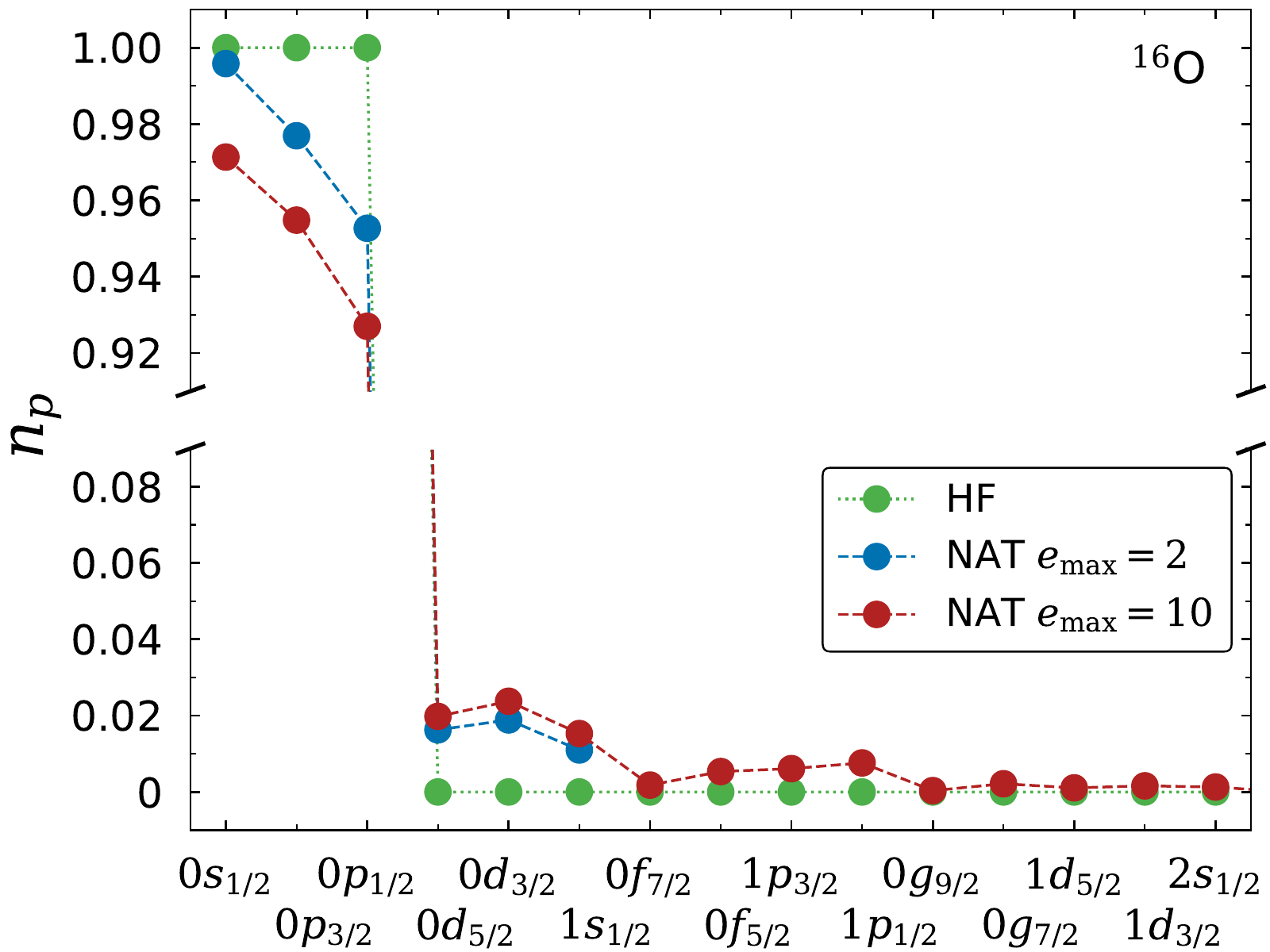}
\caption{
\label{fig:NAT_occ}
Occupation numbers $n_p$ of the single-particle proton orbitals for the HF and NAT basis in $^{16}$O, using the 1.8/2.0 EM interaction~\cite{Hebe11fits} and an oscillator frequency \hw $=16$~MeV.
We show results for two model-space truncations $e_\text{max}= 2$ and $e_\text{max}=10$ in the NAT basis construction. As for Fig.~\ref{fig:rho_MP2_contour}, the occupations of the neutron orbitals are nearly identical.}
\end{figure}

Note that the set of occupation numbers for the natural orbitals $n_p\in [0,1]$ obtained from the eigenvalues now leads to a fractional filling of all orbitals, in contrast to the occupation numbers $n_p\in \{0,1\}$ obtained from the HF solution.
This feature is illustrated in Fig.~\ref{fig:NAT_occ} comparing the NAT and HF occupation numbers for an $^{16}$O reference state.

Since the reference state for the MP2 density matrix is not a single Slater determinant due to mixing of particle-hole excitations the occupation numbers must differ from the mean-field picture.
As discussed in the following, this also affects the normal-ordering procedure with respect to natural-orbital basis states.

While the employed MP2 density matrix provides a simple approximation to the exact one-body density matrix, nonperturbative many-body schemes can be used to refine the approximation, e.g., a $\Lambda$ approach in CC theory~\cite{Shav09MBmethod}, dressed propagators from Green's function theory~\cite{Carb13SCGF3B}, or a fully correlated configuration interaction (CI) calculation~\cite{Const17NatHe6}.
A balance between accuracy and computational complexity needs to be found, and a low-order MBPT approach provides a reasonable approximation to the one-body density matrix at low computational cost.

\subsection{Intrinsic kinetic energy}
\label{sec:tkin}

The intrinsic Hamiltonian, here considered up to three-body contributions, can be split into a kinetic part and an interaction part,
\begin{align}
H = T- T_\text{cm} + V^{(2)} + V^{(3)} =  T_\text{int} + V^{(2)} + V^{(3)} \,,
\end{align}
with the intrinsic kinetic energy $T_{\text{int}}$ and the two- and three-body potentials $V^{(2)}$ and $V^{(3)}$, respectively.
The intrinsic kinetic energy is obtained by subtracting the center-of-mass kinetic energy $T_\text{cm}$ from the full kinetic energy $T$.
The intrinsic kinetic energy can be represented either as a sum of one- and two-body operators,
\begin{align}
\label{eq:tint_1+2}
T_\text{int}^{(1+2)} = \left(1-\frac{1}{A} \right) \sum \limits_i \frac{\textbf{p}_i^2}{2 m} - \frac{1}{A} \sum \limits_{i<j} \frac{\textbf{p}_i \cdot \textbf{p}_j}{m} \,,
\end{align}
or as a pure two-body operator,
\begin{align}
\label{eq:tint_2}
T_\text{int}^{(2)} = \frac{1}{A} \sum \limits_{i<j} \frac{(\textbf{p}_i-\textbf{p}_j)^2}{2m} \,.
\end{align}
Of course, both cases are equal and can be transformed into each other by
\begin{align}
\begin{split}
\sum \limits_{i<j} \frac{(\textbf{p}_i-\textbf{p}_j)^2}{2m} &= \sum \limits_{i<j} \frac{(\textbf{p}_i^2 + \textbf{p}_j^2 - 2 \textbf{p}_i \cdot \textbf{p}_j )}{2m} \\
&= (A-1) \sum \limits_{i} \frac{\textbf{p}_i^2}{2m} - \sum \limits_{i<j} \frac{\textbf{p}_i \cdot \textbf{p}_j }{m} \,.
\end{split}
\end{align}
The one- and two-body matrix elements of the Hamiltonian obviously differ depending on the choice of $T_\text{int}$.
Nevertheless both cases result in the same HF determinant with identical total HF energy, as studied in detail in Refs.~\cite{Khad74TkinHF,Jaqu92TkinSM}. The HF single-particle energies are different for both choices and can be related by a unitary transformation of the occupied single-particle states~\cite{Khad74TkinHF}.
These findings are based on the assumption of a reference state with well-defined particle number $A$. For a discussion of particle-number breaking theories, e.g., the Hartree-Fock-Bogoliubov approach, see Ref.~\cite{Herg09TkinHFB}.

The partitioning of the kinetic energy operator also affects the construction of the natural orbital basis. By employing $T_\text{int}^{(2)}$ the initial Hamiltonian (before normal ordering) no longer has a one-body part,
and the two-body matrix elements in the construction of $\gamma^\text{MP2}$ [see Eqs.~\eqref{eq:DAhp}--\eqref{eq:DDpp}] differ from the ones obtained by using the one- plus two-body form of the kinetic energy, resulting in altered transformation coefficients and NAT occupation numbers.
The partitioning also changes the single-particle energies, further changing the resulting $\gamma^\text{MP2}$.

In general, we apply the intrinsic kinetic energy operator of Eq.~\eqref{eq:tint_1+2} with a one- and two-body part for the IMSRG calculations performed in this work. However, in the following we additionally study the impact of using a pure two-body kinetic energy operator $T_\text{int}^{(2)}$.

\section{Normal ordering}
\label{sec:normalordering}

The concept of normal ordering facilitates the formulation of Wick's theorem~\cite{Wick50theorem} and defines an in-medium optimized representation of the operator~\cite{Bogn10PPNP}. For this reason, it is commonly employed in many-body frameworks applied to nuclei and nuclear matter. In the following, we first address the formal details of working with a correlated reference state.

\subsection{Multi-reference formulation}
\label{sec:mr_NO}

When employing a perturbatively improved one-body density matrix the associated reference state is no longer a single Slater determinant.
Consequently, the notion of normal ordering needs to be extended to cope with the multiconfigurational character of the vacuum. Such an extension can be naturally addressed in terms of the generalized Mukherjee-Kutzelnigg normal ordering~\cite{Kutz97mrNO}.
Even though this scheme is not numerically benchmarked in this work, it is still worth anticipating the additional complications that arise from a multireference treatment of the MP2 density matrix.

For simplicity, we neglect three-body contributions in the
following analysis and start from an arbitrary many-body operator $O$ containing up to two-body contributions:
\begin{align}
O \equiv \op{O}{0}{} + \op{O}{1}{} + \op{O}{2}{}\, .
\label{eq:generalop}
\end{align}
Performing the normal ordering of the operator $O$ in Eq.~\eqref{eq:generalop} with respect to a non-product-type vacuum yields~\cite{Gebr16MR}
\begin{subequations}
\begin{align}
\opno{o}{0}{} &= \op{o}{0}{} + \sum_{pq} \op{o}{1}{pq} \gamma_{pq} + \frac{1}{4} \sum_{pqrs} \op{o}{2}{pqrs} \gamma_{pqrs} \, , \label{eq:opnomultiref_a} \\
\opno{o}{1}{pq} &= \op{o}{1}{pq} + \sum_{rs} \op{o}{2}{prqs} \gamma_{rs}  \, , \label{eq:opnomultiref_b}\\
\opno{o}{2}{pqrs} &= \op{o}{2}{pqrs}  \,, \label{eq:opnomultiref_c}
\end{align}%
\label{eq:opnomultiref}%
\end{subequations}
involving one- and two-body density matrices $\gamma_{pq}$ and $\gamma_{pqrs}$, respectively.
The two-body density matrix, which contributes to the zero-body part of the normal-ordered operator, is given by
\begin{align}
\gamma_{pqrs} \equiv \frac{ \la \Psi | : a^\dagger_p a^\dagger_q a_s a_r :| \Psi \ra}{\la \Psi | \Psi \ra}
\end{align}
and can be decomposed into a factorized part of products of one-body operators and an irreducible two-body part $\lambda_{pqrs}$,
\begin{align}
\gamma_{pqrs} =  \Big ( \gamma_{pr} \gamma_{qs} - \gamma_{ps} \gamma_{qr} \Big ) + \lambda_{pqrs}   \, .
\end{align}
The appearance of $\lambda_{pqrs}$ is a consequence of the reference state being no longer of mean-field character.\footnote{In practice, such states are obtained, e.g., from particle-number-broken and -restored Hartree-Fock-Bogoliubov vacua~\cite{Herg13MR} or small-scale CI diagonalizations~\cite{Gebr17IMNCSM}.}
In the following, the irreducible two-body part is discarded for simplicity and a mean-field-like approximation is employed:
\begin{align}
\gamma_{pqrs} \approx  \gamma_{pr} \gamma_{qs} - \gamma_{ps} \gamma_{qr}  \, .
\label{eq:2rdmHF}
\end{align}
Equation~\eqref{eq:2rdmHF} is exact as long as a many-body state of product type is used.\footnote{Approximating a two-body density matrix from one-body density matrices is closely related to the approach followed in SCGF theory, where higher-order Green's functions are typically factorized products of one-body Green's functions, thus neglecting their irreducible higher-body contributions~\cite{Carb13SCGF3B}.}
While in principle it is straightforward to derive two-body density matrices in MBPT, the factorized approximation is expected to provide a reasonably good choice for basis optimization.

Considerable simplifications are obtained by working in the natural orbital basis, i.e.,
\begin{align}
\gamma_{pq} = n_p \delta_{pq}\,, \quad \quad n_p \in \left [0,1 \right ],
\end{align}
where the lack of a well-defined particle-hole picture means the occupation numbers are no longer zero or one. 
Expressions for the normal-ordered matrix elements in Eq.~\eqref{eq:opno} are
\begin{subequations}
\begin{align}
\label{eq:opnonator_0b}
\opno{o}{0}{} &= \op{o}{0}{} + \sum_{p} \op{o}{1}{pp} n_p + \frac{1}{4} \sum_{pq} \op{o}{2}{pqpq} n_p n_q \, , \\
\label{eq:opnonator_1b}
\opno{o}{1}{pq} &= \op{o}{1}{pq} + \sum_{r} \op{o}{2}{prqr} n_r   \, , \\
\label{eq:opnonator_2b}
\opno{o}{2}{pqrs} &= \op{o}{2}{pqrs}  \,,
\end{align}%
\label{eq:opnonator}%
\end{subequations}
now involving single-particle summations running over the full one-body Hilbert space for the summation indices $p$, $q$, and $r$ instead of hole orbitals only, a consequence of the smeared-out Fermi distributions in the occupation numbers $n_p$, as shown in Fig.~\ref{fig:NAT_occ}.

\subsection{Single-reference case}
\label{sec:srno}

In the simplest case, a single Slater-determinant reference state is employed in the many-body expansion,
\begin{align}
| \Phi \ra = \prod_{i=1}^A a^\dagger_i | 0 \ra \,,
\label{eq:prodref}
\end{align}
where $\{a^\dagger_i\}$ denotes the single-particle creation operators in the computational basis.
For the occupation numbers of the individual orbitals one has
\begin{align}
n_p =
\begin{cases}
1 \quad\text{if $p$ is a hole state}\\
0 \quad \text{if $p$ is a particle state} \,.
\end{cases} 
\label{eq:occnum}
\end{align}
Performing the single-reference normal ordering with respect to the reference state in Eq.~\eqref{eq:prodref}, the corresponding normal-ordered matrix elements of the operator are obtained as~\cite{Bogn10PPNP}
\begin{subequations}
\begin{align}
\label{eq:opno_0b}
\opno{o}{0}{} &= \op{o}{0}{} + \sum_{i} \op{o}{1}{ii} + \frac{1}{2} \sum_{ij} \op{o}{2}{ijij} + \frac{1}{6} \sum_{ijk} \op{o}{3}{ijkijk} \, , \\
\label{eq:opno_1b}
\opno{o}{1}{pq} &= \op{o}{1}{pq} + \sum_{i} \op{o}{2}{piqi} + \frac{1}{2} \sum_{ij} \op{o}{3}{pijqij}   \, , \\
\label{eq:opno_2b}
\opno{o}{2}{pqrs} &= \op{o}{2}{pqrs} + \sum_{i} \op{o}{3}{pqirsi}  \,, \\
\label{eq:opno_3b}
\opno{o}{3}{pqrstu} &= \op{o}{3}{pqrstu} \,, 
\end{align}
\label{eq:opno}%
\end{subequations}
where the labels $i$, $j$ indicate hole states occupied in the reference state $|\Phi\ra$.
In Eqs.~\eqref{eq:opno_0b}--\eqref{eq:opno_3b} three-body contributions are explicitly included. In practice, the normal-ordered two-body (NO2B) approximation is employed~\cite{Hage07CC3N,Roth12NCSMCC3N}, where the residual three-body part, Eq.~\eqref{eq:opno_3b}, is discarded to lower the computational complexity.

Because the MP2 density matrix does not correspond to a single Slater determinant, an auxiliary many-body state $| \NAT \ra$ is constructed by filling the first $A$ states with the highest occupation numbers.
Similar to Eq.~\eqref{eq:prodref} these orbitals are filled with updated occupations $n_i \in \{0,1\}$ to conserve the particle-number
expectation value, thus establishing a well-defined particle-hole picture.
Consequently, in the following applications standard Slater-determinant-based codes can be used for the many-body expansion.
Note that, even though this reference state has product-type character, the information about the correlated density matrix is encoded in the transformation matrix from the HO to the NAT basis [see Eq.~\eqref{eq:CNatHF}] for the one- and two-body parts of the intrinsic Hamiltonian.
By using such an auxiliary vacuum the reference-state expectation value is larger than the HF expectation value since there is no underlying variational principle, i.e.,
\begin{align}
\label{eq:Eexpval}
\la \NAT | H | \NAT \ra
> \la \HF | H | \HF\ra \, .
\end{align}

\section{In-medium similarity renormalization group}
\label{sec:imsrg}

For the medium-mass applications in this work we use the nonperturbative in-medium similarity renormalization group approach.
For a detailed discussion of the many-body formalism the reader is referred to Refs.~\cite{Tsuk11IMSRG,Herg16PR,Herg17PS}.

\subsection{Formalism}

In the IMSRG framework  the many-body Schr\"odinger equation is solved by performing a decoupling of particle-hole excitations from the reference state by a continuous unitary transformation $U(s)$ parametrized in terms of a real-valued flow parameter $s$,
\begin{align}
H(s) = U(s) H_0 U^\dagger(s) \, ,
\label{eq:imsrg}
\end{align}
where $H(s=0)$ is the initial, i.e., unevolved, Hamiltonian.
Equation~\eqref{eq:imsrg} can be rewritten as a first-order ordinary differential equation (ODE) in $s$,
\begin{align}
\frac{\mathrm{d}H(s)}{\mathrm{d}s} = \left[\eta(s), H(s) \right] \,,
\label{eq:imsrgflow}
\end{align}
with the anti-Hermitian generator $\eta(s)$ defined by the unitary transformation.

The in-medium character of the decoupling condition is achieved by performing the renormalization group evolution on the normal-ordered representation of the operator.
In the simplest case this is done with respect to a single Slater determinant, as indicated in Sec.~\ref{sec:srno}.
At $s=0$, the normal-ordered zero-body part of the Hamiltonian is the reference-state energy expectation value, e.g., the HF energy in the case of the HF reference state.
Over the course of the evolution,
the off-diagonal matrix elements of the Hamiltonian are suppressed,
and the exact ground-state energy of the fully interacting system
is given by the normal-ordered zero-body part of the evolved Hamiltonian
\begin{align}
E_0(s\rightarrow \infty) = \lim_{s\rightarrow \infty} \la \Phi | H(s) | \Phi \ra \, .
\end{align}
Solving the IMSRG flow equation absorbs the dynamic correlations nonperturbatively and smoothly decouples particle-hole excitations from the reference state.

In practice, the expansion in Eq.~\eqref{eq:imsrgflow} is computationally intractable since the repeated commutator evaluation induces higher-body operators in each integration step, yielding up to $A$-body operators when applied to an $A$-body system.
In the following, the IMSRG(2) approximation is used, where all operators are truncated at the normal-ordered two-body level.
The IMSRG(2) approximation provides an accurate truncation scheme that incorporates all MBPT corrections up to order $\lambda^3$
while resumming many higher-order contributions in a nonperturbative way through the IMSRG flow~\cite{Herg16PR}.
In medium-mass applications, the many-body uncertainty related to the IMSRG(2) approximation is estimated to be $\approx 2\%$ for ground-state energies (see, e.g., Ref.~\cite{Huth19chiralfam}), thus providing a versatile and precise many-body scheme at moderate computational cost.

In all subsequent calculations the modified normal ordering as discussed in Sec.~\ref{sec:srno} is employed, thus enabling the use of the standard Slater-determinant-based IMSRG.
Even though various open-shell extensions have been designed and applied within the IMSRG approach~\cite{Herg13MR,Stro17ENO,Gebr17IMNCSM}, this work is restricted to closed-shell applications to provide a simple testbed for the natural orbital basis.

\subsection{Magnus reformulation}
\label{sec:magnus}
In this work, the Magnus formulation of the IMSRG~\cite{Magn54exp,Morr15Magnus} is employed, which provides direct access to the unitary transformation $U(s)$ in a computationally efficient way using the parametrization
\begin{align}
U(s) = e^{\Omega(s)} \, ,
\end{align}
where $\Omega(s)$ is the anti-Hermitian Magnus operator.
Analogously to Eq.~\eqref{eq:imsrgflow}, an ODE for the Magnus operator can be derived.
Having the transformation matrix at hand allows for the IMSRG evolution of any other operator by the Baker-Campbell-Hausdorff formula for the many-body operator according to
\begin{align}
O(s) = e^{\Omega(s)} O(s=0) e^{-\Omega(s)}
\label{eq:bch}
\end{align}
instead of solving an additional set of ODEs for each operator.
In practice, the Baker-Campbell-Hausdorff expansion is performed using nested commutator evaluations to some truncation level.
Moreover, solving the Magnus expansion has the advantage of allowing the use of a simpler ODE solver for the Magnus operator without loss of accuracy.

\subsection{Correlation effects from MP2 natural orbitals}
\label{sec:corr_effects_from_MP2}

Before discussing IMSRG applications using natural orbitals, it is worth addressing the interplay of the correlations built into the MP2 density matrix and the correlations that are resummed within the IMSRG flow.

Using a natural orbital reference determinant yields a higher ground-state energy at $s=0$ compared to an HF vacuum due to the variational optimization of the HF orbitals in the space of single Slater-determinant reference states [see Eq.~\eqref{eq:Eexpval}].
Moreover, the ground-state energy at $s=\infty$ using the MP2 density matrix does not improve upon the IMSRG(2) results obtained in any other single-particle basis.
The MP2 density matrix only incorporates correlations to one-particle-one-hole and two-particle-two-hole excitations.
Within the IMSRG(2) approximation such effects are resummed to all orders~\cite{Herg16PR} such that no improvement on the final observable is expected.
Once higher-body excitations are included, additional correlations will enter the description which are absent in the IMSRG(2) scheme. 
Practically, this is achieved by including third-order terms in the MBPT expansion, i.e., $\lambda^3$, or allowing three-body operators in the normal-ordered Hamiltonian, thus generating additional contributions in the first-order state correction.
Both options will generate the leading contributions to three-particle-three-hole excitations.

\section{Diagnostics for the density matrix}
\label{sec:results1}

We begin by investigating the MP2 density matrix and the associated NAT basis. 
These results allow us to gain a better understanding of the relationship between the various computational bases and their sensitivity to the nuclear Hamiltonian used in their construction. 
We focus on two sets of chiral interactions, the N$^3$LO \textit{NN} potential from Ref.~\cite{Ente17EMn4lo} with N$^3$LO 3\textit{N} forces constructed in Ref.~\cite{Dris17MCshort}, which in the following is referred to as ``N$^3$LO'' with the corresponding cutoff value, and the ``1.8/2.0 EM'' interaction of Ref.~\cite{Hebe11fits}.

\subsection{The ``softness'' of the interaction}
\label{sec:soft_int}

``Soft'' interactions are low-resolution interactions that show weak coupling between low- and high-energy states.
The softness of an interaction, i.e., the degree of decoupling between low and high momenta in the Hamiltonian, can be varied by changing the regulator scale
for Hamiltonians constructed from an EFT
as well as by applying (similarity) renormalization group [(S)RG] methods to decouple or integrate out high-momentum degrees of freedom \cite{Bogn10PPNP,Furn12NPPS}.
Soft interactions applied in many-body methods have been shown to improve convergence with respect to basis truncation and order in the many-body expansion.
In particular, the use of an SRG-evolved Hamiltonian is required to enable a perturbative solution even for closed-shell systems~\cite{Tich16HFMBPT}.
A Weinberg-eigenvalue analysis, which provides a metric of the perturbativeness of an interaction,
shows that the softness is intimately linked to the SRG resolution scale~\cite{Bogn06bseries,Hopp17WeinEVAn}.

Because the one-body density matrix is constructed from an MBPT expansion, we expect the density matrix and the resulting NAT basis to be more sensitive to the basis frequency and truncation for hard interactions.
For unevolved chiral potentials the mean-field wave function may exhibit unphysical properties, giving rise to an unbound HF solution.
With such a poor reference state, the many-body expansion is significantly more complicated, in particular if perturbative techniques are employed.
The key idea of a many-body expansion is to start from a qualitatively correct reference state while residual dynamic correlation effects are brought in as (small) corrections.
This rationale is obviously broken once the mean-field reference is unbound or not under control, manifesting in final results via, e.g., strong frequency dependence.

\begin{figure}[t!]
\centering
\includegraphics[width=\columnwidth,clip=]{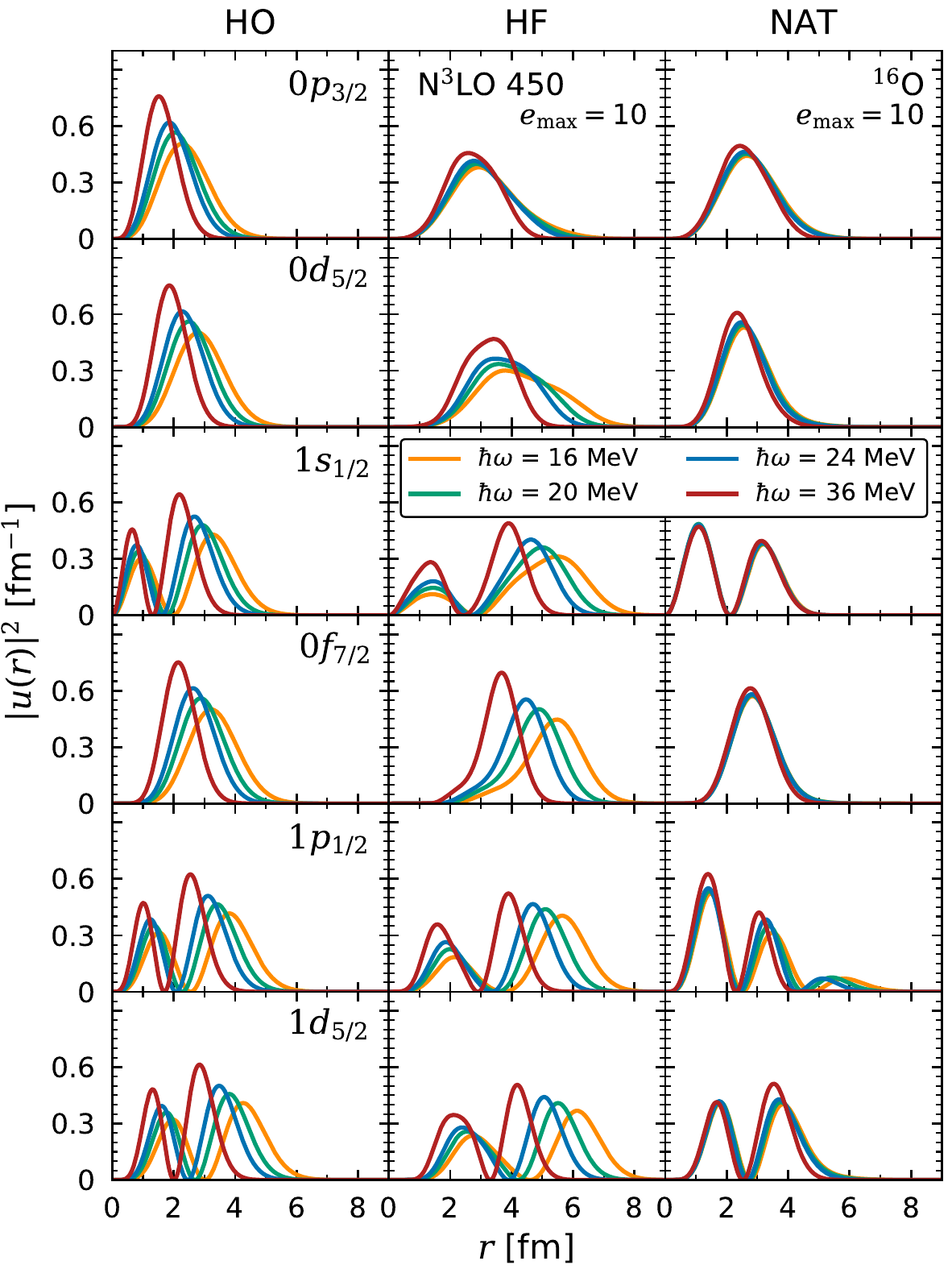}
\caption{
\label{fig:radial_wf_emax10_HF+NAT}
Squared absolute value of the radial wave function $u(r)$ of $^{16}$O as a function of $r$ for different proton orbitals in the HO, HF, and NAT bases in the first, second, and third columns, respectively.
We show results for the occupied $0p_{3/2}$ (first row) and some of the first unoccupied orbitals (second through sixth rows) for the N$^3$LO 450 interaction and oscillator frequencies \hw $= 16$--$36$~MeV.
The HF and NAT orbitals include single-particle HO states up to $e_\text{max}=10$ and $E_\text{3max} = 14$.}
\end{figure}

\subsection{Single-particle wave functions}
\label{sec:sp_wave_func}

While the HF approach targets the optimization of the occupied single-particle states from a variational approach, the unoccupied orbitals are left unmodified up to normalization. 
Therefore, the HF basis is expected to properly describe occupied orbitals while failing for unoccupied ones.
The natural orbital basis, however, accounts for particle-hole admixtures and therefore may qualitatively improve the description of unoccupied states as will be tested in the following calculations.
In the following, a single-particle basis is employed including states up to a principal quantum number $e_\text{max}$.
Additionally, we introduce a truncation in three-body space keeping only configurations with $e_1 + e_2 + e_3 \leq E_\text{3max} < 3e_\text{max}$ due to the extensive size of three-body matrix elements.

In Fig.~\ref{fig:radial_wf_emax10_HF+NAT}, we show the squared absolute value of the radial wave functions for different oscillator frequencies using the HO, HF, and NAT bases. 
Different rows correspond to different single-particle orbitals; only the first row $(0p_{3/2})$ corresponds to an occupied orbital.
Clearly, using a HO basis leads to strong frequency dependence in all cases, even for the occupied $0p_{3/2}$ state.
Hence, HO wave functions are ruled out as a reliable computational basis and are not considered further in this work.
While the $0p_{3/2}$ orbitals are more robust in the HF case as expected, unoccupied HF orbitals show frequency dependence comparable to that of HO orbitals, a consequence of the fact that unoccupied orbitals are not optimized in the HF approach.

\begin{figure}[t!]
\centering
\includegraphics[width=\columnwidth,clip=]{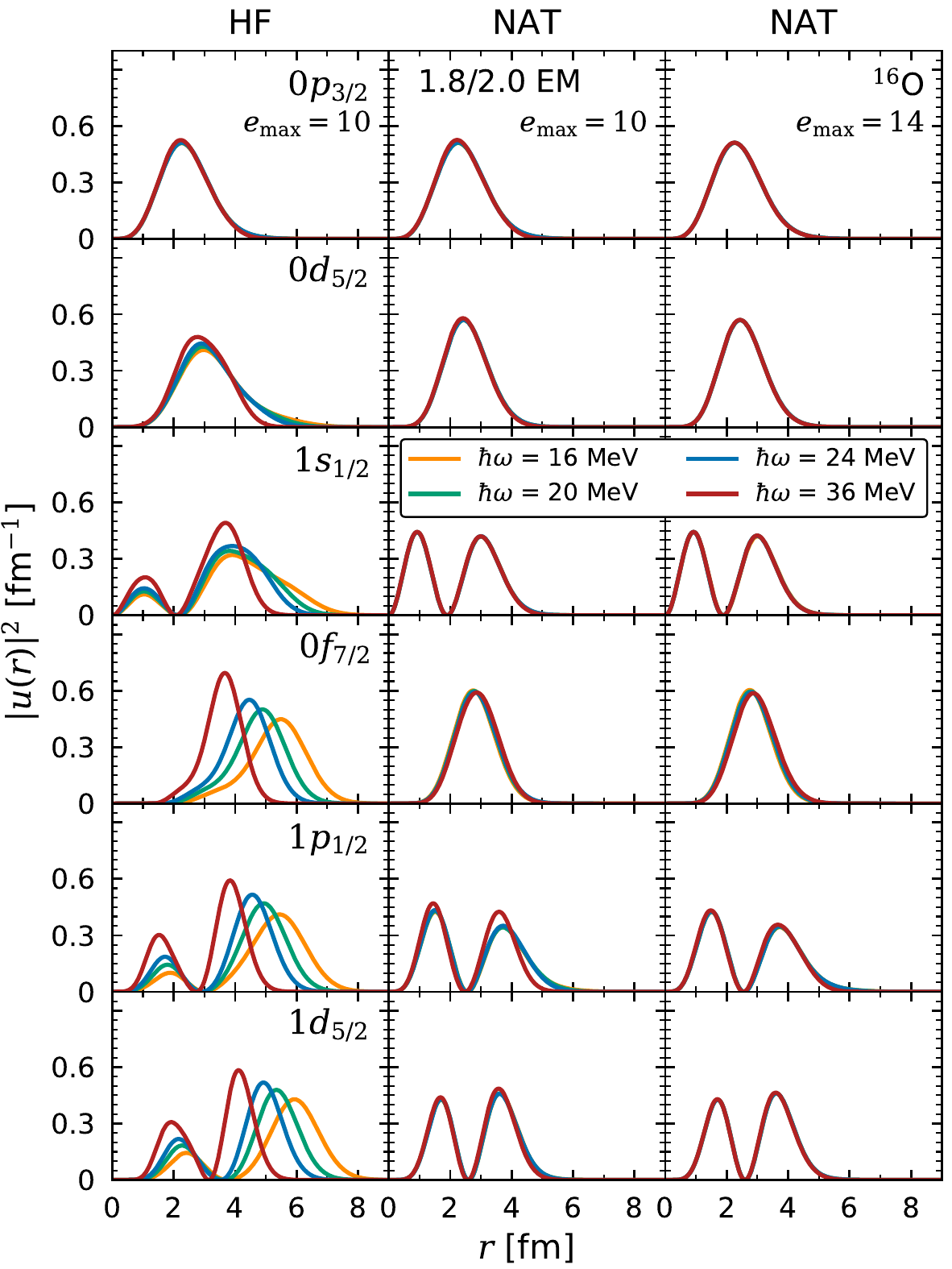}
\caption{
\label{fig:radial_wf_NAT_E2}
Same as Fig.~\ref{fig:radial_wf_emax10_HF+NAT} but for the 1.8/2.0 EM interaction showing results for the HF and NAT bases using $e_\text{max}=10$ in the left and middle columns, respectively, as well as $e_\text{max}=14$ for the NAT basis in the right column.}
\end{figure}

Switching to natural orbitals nicely resolves many of the remaining artifacts, revealing only minor frequency dependence for both occupied and unoccupied states.
As the softer 1.8/2.0 EM interaction from Ref.~\cite{Hebe11fits} leads to much better reproduction of ground-state energies at the HF level~\cite{Simo17SatFinNuc}, this also improves the quality of the MP2 density matrix.
In Fig.~\ref{fig:radial_wf_NAT_E2}, we compare HF (left) and natural orbitals (middle) in a model space with $e_\text{max}=10$ for this interaction, while additionally benchmarking the effect of natural orbitals when going to a larger basis size of $e_\text{max}=14$ (right). While the frequency dependence for this softer interaction is much milder in the HF case, high-lying single-particle states still significantly depend on $\hbar \omega$. A residual frequency dependence is still seen in the $1p_{1/2}$ orbital at $e_{\text{max}}=10$ in the natural orbital basis, but this fully vanishes when going to larger spaces of $e_\text{max}=14$.

In summary, properties of the HF solution strongly impact the qualitative behavior of the natural orbital single-particle wave functions and a bound mean-field solution is key for providing a reliable reference point for a many-body expansion.

\begin{figure}[t!]
\centering
\includegraphics[width=\columnwidth,clip=]{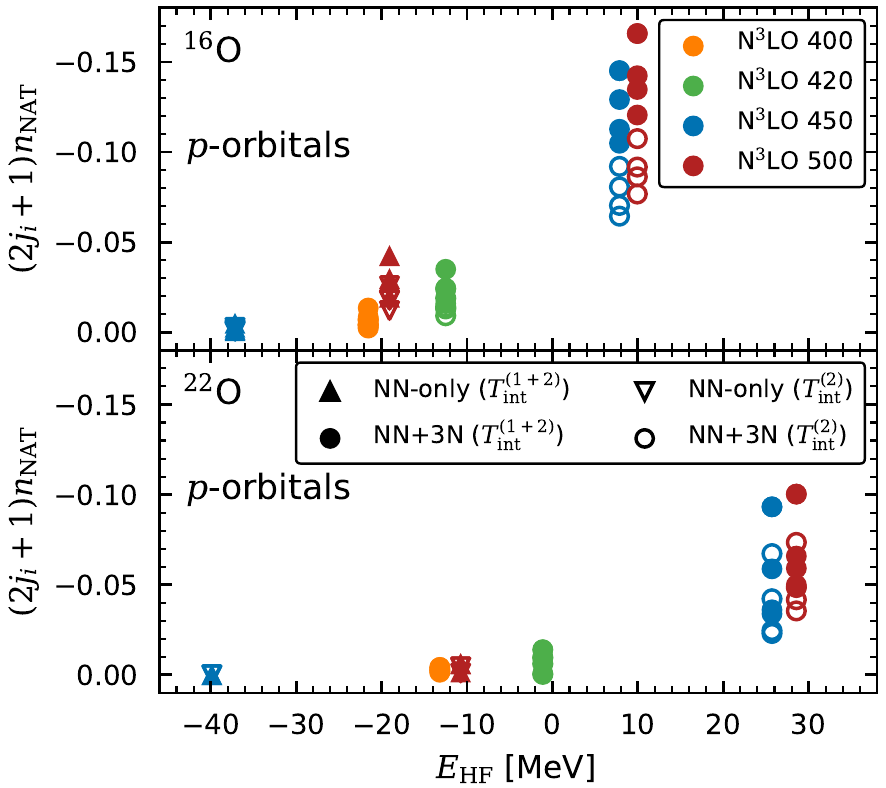}
\caption{
\label{fig:neg_occ_C12_O16_O22}
Negative occupations of the $p$ orbitals scaled by $(2j_i+1)$ in the NAT basis for $^{16}$O (top) and $^{22}$O (bottom) as a function of the HF energy. We show results for various cutoffs with the \textit{NN}-only N$^3$LO EMN and \textit{NN+{\normalfont 3}N} N$^3$LO interactions indicated by triangles and circles, respectively. We apply both choices for the kinetic energy operator, $T_\text{int}^{(1+2)}$ (solid symbols) and $T_\text{int}^{(2)}$ (open symbols), and use a model space of $e_\text{max}/E_{3\text{max}} = 14/14$ with \hw$=20$~MeV.
All negative occupations arise only for high radial quantum number. Note that there are no negative occupations for the softer \textit{NN}-only EMN 400 and 420 interactions.}
\end{figure}

\begin{figure*}[t!]
\centering
\begin{minipage}[c]{.49\textwidth}
\centering
\includegraphics[width=\columnwidth,clip=]{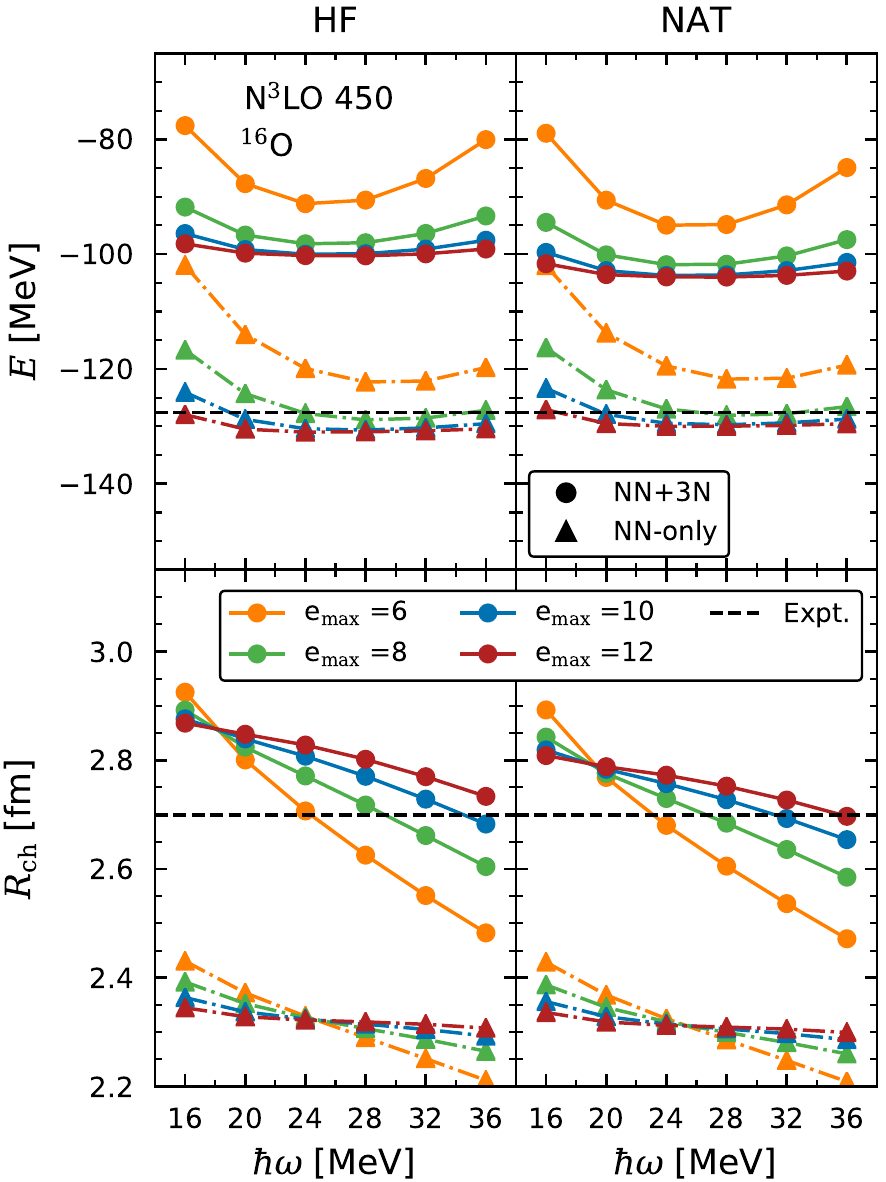}
\end{minipage}
\hspace*{0.16cm}
\begin{minipage}[c]{.49\textwidth}
\centering
\includegraphics[width=\columnwidth,clip=]{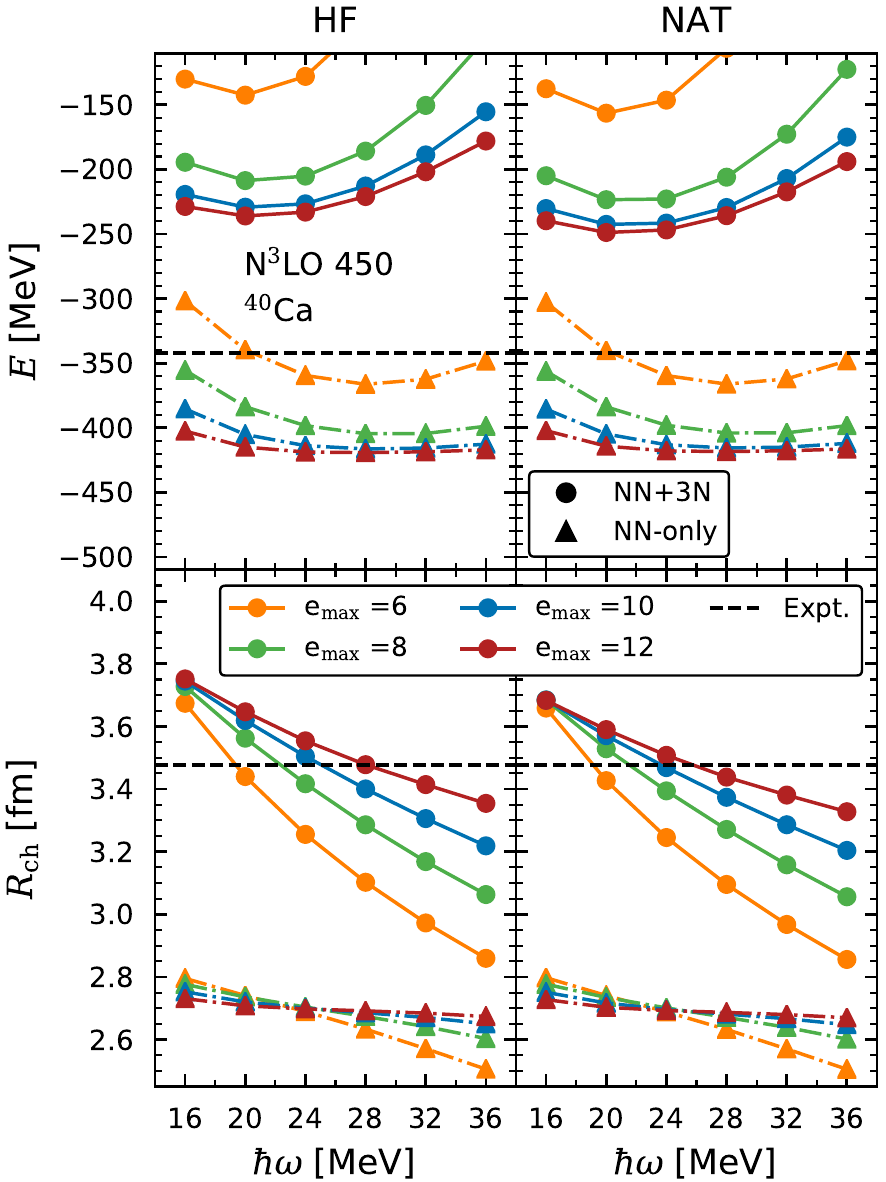}
\end{minipage}
\caption{
\label{fig:HF_NAT_O16_Ca40_EMN450}
Ground-state energies (upper rows) and charge radii (lower rows) of $^{16}$O and $^{40}$Ca in the left and right plots, respectively, as a function of the oscillator frequency, for the \textit{NN}-only N$^3$LO EMN 450 (triangles) and \textit{NN+{\normalfont 3}N} N$^3$LO 450 (circles) interactions. We show results for the HF and NAT bases in the left and right panels of each plot, respectively, using various single-particle truncations $e_\text{max}$ with $E_{\text{3max}}=14$. Experimental values are taken from Refs.~\cite{Ange13rch,Wang17AME16}.}
\end{figure*}

\subsection{Positive definiteness as diagnostic tool}

The density matrix is a positive-definite operator and thus its eigenvalues, the occupation numbers, are non-negative.
Therefore, \textit{unphysical} negative occupations or occupations larger than one should not show up during the diagonalization.
Previous investigations in quantum chemistry showed that the appearance of negative occupation numbers can be linked to a breakdown of a single-reference description and hint at the onset of strong static correlations~\cite{Gord99negNAT}.
Therefore, we aim to utilize occupation numbers as diagnostic and investigate their sensitivity to the softness of the nuclear interaction.
As the HF ground-state energy is directly related to the softness of the interaction, a correlation between the HF energy and the size of negative occupations is expected.

Figure~\ref{fig:neg_occ_C12_O16_O22} depicts the magnitude of the negative occupations using N$^3$LO interactions for various cutoff values in $^{16,22}$O.
In both nuclei, we observe a decrease in size for softer interactions, as indicated by going from the harder potentials with cutoff $\Lambda=500$~MeV to $\Lambda=400$~MeV, in both the \textit{NN}-only and the \textit{NN+{\normalfont 3}N} cases.
Consequently, an unbound HF solution strongly affects the appearance of unphysical negative occupations. 
In general, using the two-body form of the kinetic energy operator $T_\text{kin}^{(2)}$ results in smaller negative occupations for both nuclei.
We also verified that softening the interaction by a consistent SRG evolution of \textit{NN} and 3\textit{N} contributions~\cite{Hebe12msSRG,Hopp19medmass} significantly reduces the magnitude of the negative occupations, eventually letting them vanish completely.
Increasing the model-space size seems to increase the magnitude of these occupations. Moreover, the effect is generally less pronounced for heavier nuclei, e.g., in $^{78}$Ni.

In addition, we investigate the size of negative occupations in the case of $^{12}$C.
Due to the cluster structures and weak shell closure in $^{12}$C the quality of single-reference many-body approaches is expected to deteriorate in comparison to the doubly magic nucleus $^{16}$O.
An analysis of the single-particle spectrum revealed only a small shell gap in the single-particle spectrum, thus significantly enhancing the size of perturbative corrections to the MP2 density matrix in the particle-particle and hole-hole channel [see Eqs.~\eqref{eq:DChh}~and~\eqref{eq:DDpp}].
Consequently, highly erratic occupation numbers were observed (not shown).
Empirically, we found that the use of $T_\text{int}^{(2)}$ with a slightly larger shell gap was superior to $T_\text{int}^{(1+2)}$, significantly reducing, though not fully resolving, the large negative occupations.
The results of this analysis for the occupation numbers is also evidence for the challenges of the single reference-state starting point for a description of $^{12}$C.

\section{IMSRG results}
\label{sec:results2}

After addressing in detail properties of the single-particle basis itself, the various choices are benchmarked for medium-mass closed-shell systems using the IMSRG framework, focusing on $^{16}$O, $^{40}$Ca, and $^{78}$Ni.
All many-body calculations employed the publicly available IMSRG solver by Stroberg~\cite{Stro17imsrggit}.

\subsection{Comparing the HF and NAT basis}

We compare results for ground-state energies and charge radii of $^{16}$O and $^{40}$Ca in the HF and NAT bases in Fig.~\ref{fig:HF_NAT_O16_Ca40_EMN450} for a large range of oscillator frequencies for the \textit{NN}-only and \textit{NN+{\normalfont 3}N} N$^3$LO 450 interactions.
For the \textit{NN}-only potential, we observe nearly no change when going from the HF to the NAT basis on this scale. 
Since the HF solution is bound, bulk properties are well captured at the mean-field level and applying the NAT basis does not yield an improvement in final results.
Both energies and radii are almost flat as a function of \hw for the largest model space and rapidly converge with model-space size in both the HF and natural orbital bases.
When 3\textit{N} forces are included, the \textit{NN+{\normalfont 3}N} results similarly to the \textit{NN}-only case show almost no change from the HF to the NAT basis, but the \hw dependence becomes more pronounced for the radii.

\begin{figure}[t!]
\centering
\includegraphics[width=\columnwidth,clip=]{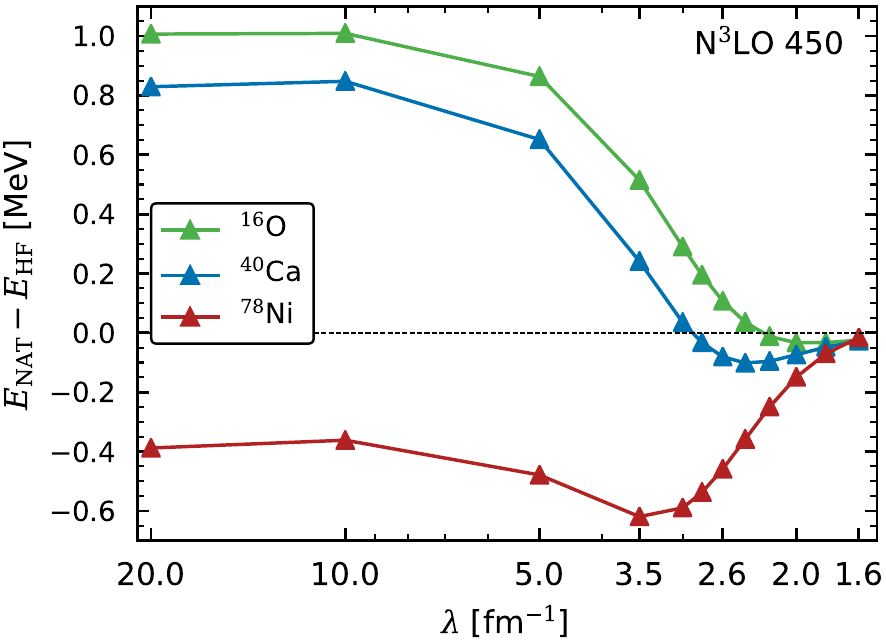}
\caption{
\label{fig:EHF_ENAT_SRG}
Difference of the ground-state energies in the NAT and HF bases for $^{16}$O, $^{40}$Ca, and $^{78}$Ni as a function of the SRG resolution scale $\lambda$ using the \textit{NN}-only N$^3$LO EMN 450 interaction and a model space of $e_\text{max}=14$ with \hw$=20$~MeV.}
\end{figure}

In order to systematically understand the difference between the basis sets, we examine the converged IMSRG(2) ground-state energies in greater detail.
In Fig.~\ref{fig:EHF_ENAT_SRG}, we show the difference of the results in the HF and NAT bases as a function of the SRG evolution scale for three closed-shell nuclei $^{16}$O, $^{40}$Ca, and $^{78}$Ni for the \textit{NN}-only interaction.
The analysis is performed in absence of three-body interactions to eliminate the sensitivity of the different reference states to the NO2B approximation.
For harder interactions (larger $\lambda$) the difference is of the order of 1 MeV with the natural orbitals yielding stronger binding for $^{16}$O and $^{40}$Ca and slightly weaker binding for $^{78}$Ni.
Softening the potential (small $\lambda$) significantly reduces the effect such that, eventually, only differences of the order of tens of keV remain at $\lambda=1.6\,\text{fm}^{-1}$.
These differences are marginally enhanced when including 3\textit{N} forces; i.e., natural orbitals provide slightly more binding compared to the HF basis and lead to a minor decrease of the \hw dependence of charge radii.
We emphasize again that this \textit{NN+{\normalfont 3}N} interaction leads to an unbound HF solution, such that the total binding has to be produced by correlation effects during the IMSRG flow and the mean field provides a poor reference state as discussed in Sec.~\ref{sec:soft_int}. 
The minor differences in converged energies are assumed to be driven by induced many-body contributions that differ in HF and natural orbital bases.
A further investigation requires systematic evaluation of leading three-body contributions beyond the IMSRG(2) approximation which is beyond the scope of the present work.

In summary, we do not observe the desired independence of the oscillator frequency in the smaller model spaces, which one could have guessed from Fig.~\ref{fig:radial_wf_emax10_HF+NAT}, and do not improve on the frequency dependence in the largest model spaces shown here compared to the HF basis.

\subsection{Differences between NCSM and IMSRG}

Given the great performance of MP2 natural orbitals in NCSM results as shown in Ref.~\cite{Tich19NatNCSM}, the above results seem surprising at first since no substantial improvement over HF orbitals is obtained. The key difference between the IMSRG calculations performed in this work so far and the NCSM calculations is the model space in which the many-body solution is obtained.

In the NCSM one conveniently employs an $N_\text{max}$ truncation where only many-body configurations up to a given relative excitation level are included~\cite{Barr13PPNP}. In this case, constructing the correlated one-body density matrix in a large single-particle basis \emph{includes excitations that are absent from the NCSM configuration space} and, therefore, improves the frequency dependence. 
On the other hand, this is inherently different from an IMSRG application where both the reference-state construction and IMSRG flow typically take place \emph{in the same model space}, parametrized by $e_\text{max}$. Since the high-lying states are included already in the initial single-particle basis for the HF calculation, we cannot expect significant improvement for the simplest natural-orbital-based IMSRG calculations over HF-based IMSRG calculations. 

Consequently, the key idea in the following will be the construction of the MP2 density matrix in a large space while solving the many-body problem in a reduced space in the presence of the full-space correlations embedded into the basis transformation.

\begin{figure*}[t!]
\centering
\begin{minipage}[c]{.49\textwidth}
\centering
\includegraphics[width=\columnwidth,clip=]{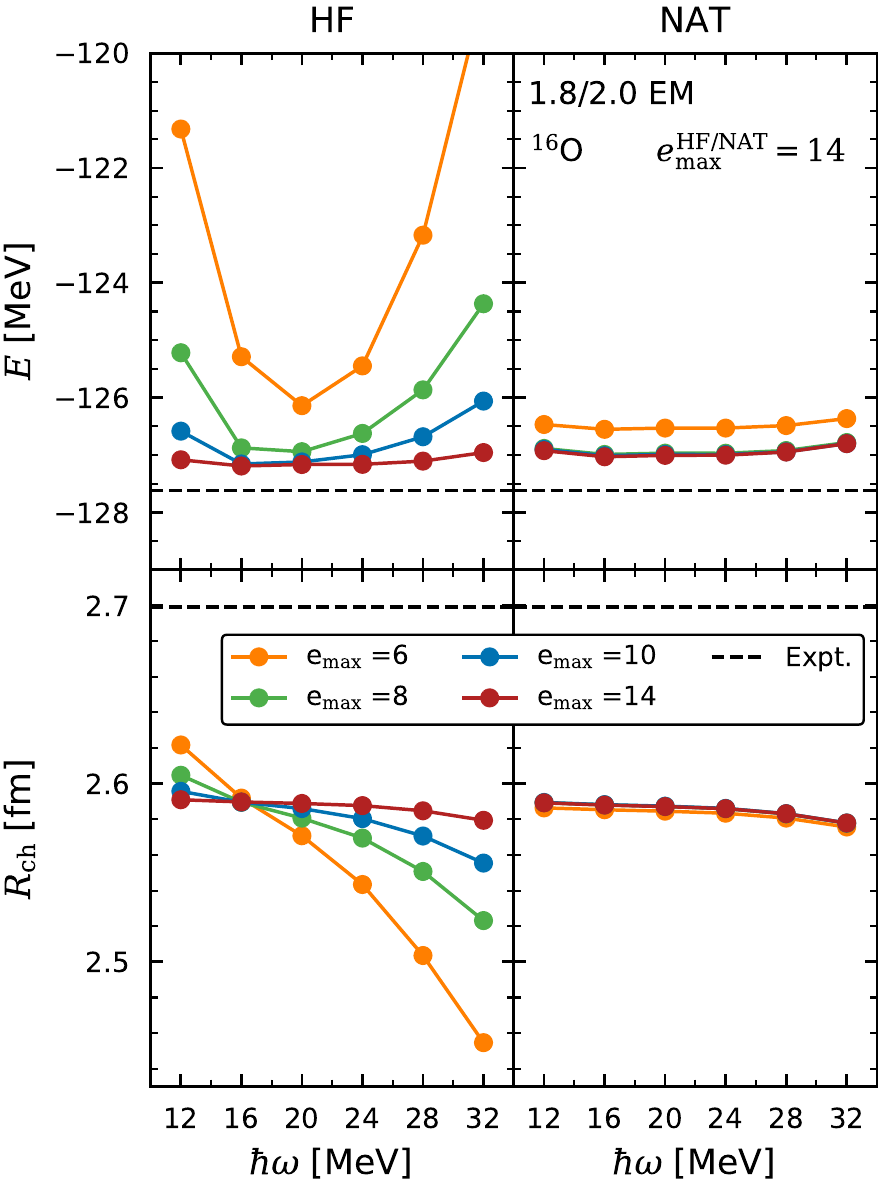}
\end{minipage}
\hspace*{0.16cm}
\begin{minipage}[c]{.49\textwidth}
\centering
\includegraphics[width=\columnwidth,clip=]{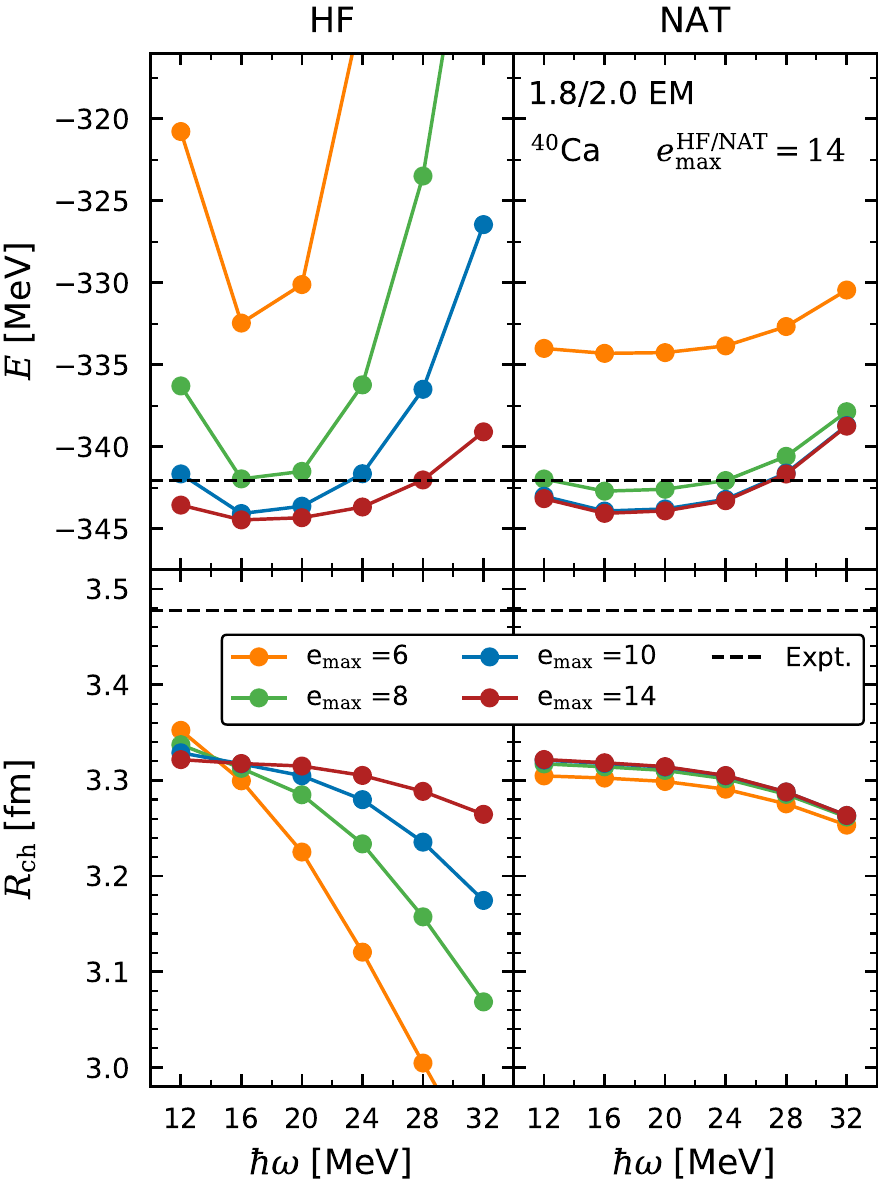}
\end{minipage}
\caption{
\label{fig:HF_NATtrunc_O16_Ca40_1.8/2.0_NN+3N}
Ground-state energies (upper rows) and charge radii (lower rows) of $^{16}$O and $^{40}$Ca in the left and right plots, respectively, as a function of the oscillator frequency in the HF and NAT bases for the 1.8/2.0 EM interaction.
We use a model space $\Mlarge$ with $e_\text{max}^\text{HF/NAT}=14$ to construct the NAT basis, whereas the IMSRG calculations are performed for $e_\text{max} = 6$, 8, 10, and 14, with $E_{3 \text{max}} = 16$ in both cases.  Experimental values are taken from Refs.~\cite{Ange13rch,Wang17AME16}.}
\end{figure*}

\begin{figure}[t!]
\centering
\includegraphics[width=\columnwidth,clip=]{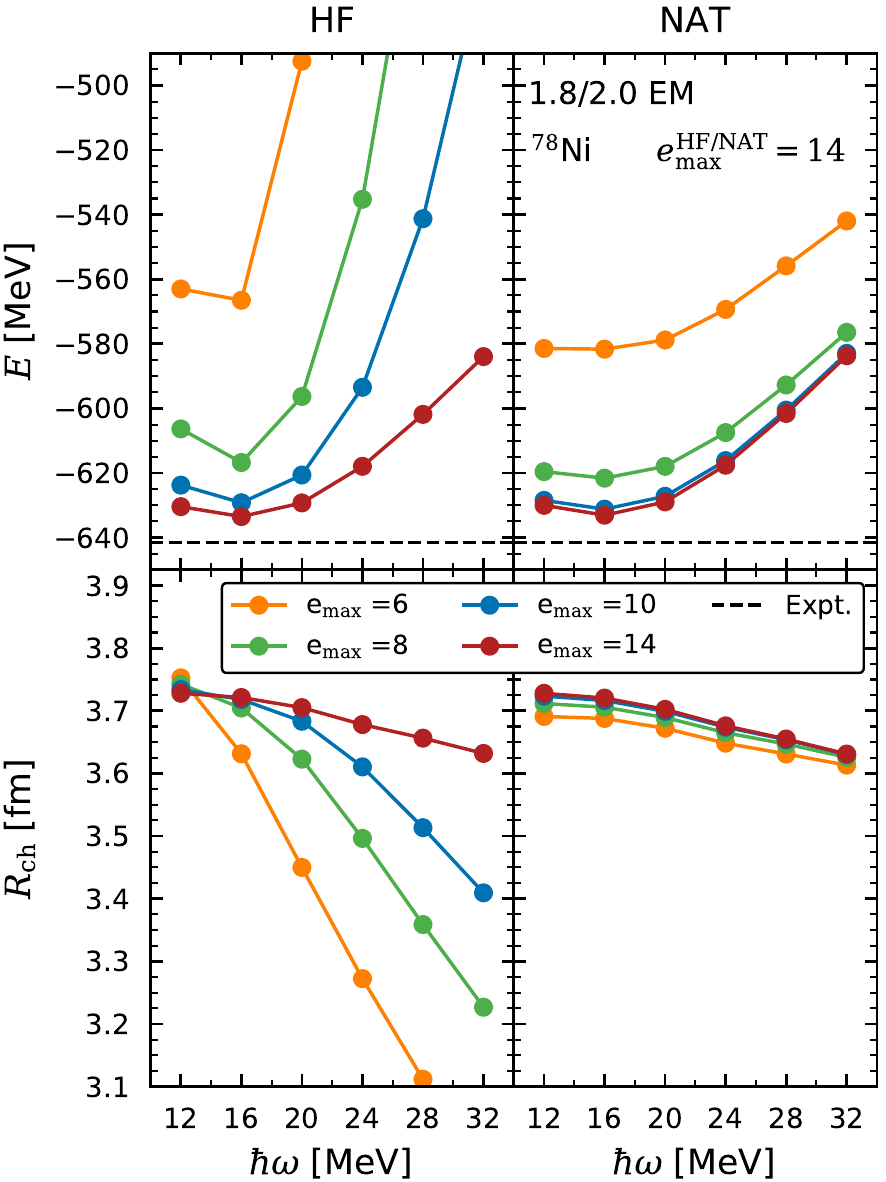}
\caption{\label{fig:HF_NATtrunc_Ni78_1.8/2.0_NN+3N}
Same as Fig.~\ref{fig:HF_NATtrunc_O16_Ca40_1.8/2.0_NN+3N} but for $^{78}$Ni. Note that the experimental ground-state energy taken from Ref.~\cite{Wang17AME16} is extrapolated.
}
\end{figure}

\subsection{Reduced-basis calculations from NAT/HF constructed in full space}

In the following, the initial MP2 density matrix is built in a large single-particle basis $\Mlarge$, while a smaller subspace $\Msmall \subseteq \Mlarge$ is used for performing the IMSRG evolution.
While the basis transformation is performed according to
\begin{align}
| n \alpha_p \ra_\text{NAT}  &=  \sum_{n^\pr} \ ^{\text{NAT/HF}}C_{n n^\pr}^{\alpha_p} | n^\pr \alpha_p  \ra_\text{HO} \,,
\label{eq:CNAT_reduced}
\end{align}
we construct a reduced basis set by keeping only a small number of the natural orbital states $e_\text{max}$ even though the density matrix construction is performed in a large space characterized by $e_\text{max}^{\text{NAT/HF}}$.
With this approach the orthonormalization of the individual basis states in the reduced space is still guaranteed.

As an example, the construction of the NAT basis states in Eq.~\eqref{eq:CNAT_reduced} in an $e_\text{max}^{\text{NAT/HF}}=10$ model space gives the optimized $s$ orbitals $0s_{1/2}$ through $5s_{1/2}$, based on the transformation of the HO states up to the $5s_{1/2}$ orbitals. A subsequent truncation to an $e_\text{max}=6$ model space discards the $4s_{1/2}$ and $5s_{1/2}$ NAT orbitals after the unitary transformation. This is to be contrasted with the construction of the $e_\text{max}=6$ NAT basis in an $e_\text{max}^{\text{NAT/HF}}=6$ model space, where there are no $4s_{1/2}$ and $5s_{1/2}$ HO orbitals present in the transformation for the natural orbital basis states. 

Even though parts of the information contained in the natural orbital basis are lost during this reduction process, the resulting matrix representation of operators in $\Msmall$ still contains information about the large space due to the mixing of radial excitations up to a maximum radial quantum number $n_\text{full}$ included in $\Mlarge$ that are otherwise not contained in $\Msmall$. As a result, this approach accounts for high radial excitations for the construction of the reduced NAT basis and leads to a better optimization of the low-lying wave functions.
Excluding higher-lying states from $\Msmall$ is also motivated by the intuition that for low-resolution Hamiltonians we expect the many-body expansion to be dominated by excitations to low-lying states.

For the following calculations, $\Msmall$ and $\Mlarge$ will be parametrized by two values, $e_\text{max}$ for the IMSRG evolution (in $\Msmall$) and $e_\text{max}^\text{HF/NAT}$ for the basis construction (in $\Mlarge$).
We employ $e_\text{max}^\text{HF/NAT}=14$ for the $1.8/2.0$ EM interaction, corresponding to the radial wave functions in the last column of Fig.~\ref{fig:radial_wf_NAT_E2} that show the desired frequency independence for this soft interaction (see Sec.~\ref{sec:soft_int}).
Comparable results are expected for the consistently SRG-evolved N$^3$LO interactions.
We investigate the impact on ground-state energies and charge radii by considering IMSRG calculations in various reduced model spaces with truncations $e_\text{max}= 6,8,$ and $10$ for $^{16}$O, $^{40}$Ca, and $^{78}$Ni in Figs.~\ref{fig:HF_NATtrunc_O16_Ca40_1.8/2.0_NN+3N} and \ref{fig:HF_NATtrunc_Ni78_1.8/2.0_NN+3N}.
Constructing the NAT basis in the full space leads to a significant reduction of the \hw dependence for both ground-state energies and charge radii as well as improved convergence behavior with respect to $e_\text{max}$.
The resulting improvement is similar to what was seen in NCSM calculations for $^{16}$O~\cite{Tich19NatNCSM} with nearly frequency-independent energies and radii, shown in the right column of the first plot in Fig.~\ref{fig:HF_NATtrunc_O16_Ca40_1.8/2.0_NN+3N}.

Analogous conclusions hold for heavier nuclei, where the convergence pattern is improved and we obtain converged results already in smaller model spaces $e_\text{max}$.
Although we cannot improve the results beyond the model space of $e_\text{max}^\text{HF/NAT}$ employed for the initial transformation, we only have to solve for the natural orbital  basis once in the largest possible $e_\text{max}^\text{HF/NAT}$ space without having to solve the computationally more expensive IMSRG equations in the full space. 
Assuming we can obtain comparable results in $e_\text{max}=10$ (1140 single-particle states) using an MP2 density matrix constructed in $e_\text{max}=14$ (2720 single-particle states), we save a factor $R \approx 2$--$3$ in single-particle dimension.
Consequently, the use of large-space natural orbitals combined with reduced-space many-body expansions provides a computationally efficient alternative to the full-space IMSRG calculations.
Very advanced truncation schemes such as IMSRG(3) scale as $N^9$, where $N$ is a measure of the size of the single-particle basis. Therefore, naive speedups of the order $R^9 \approx 10^3$--$10^4$ can be anticipated. Consequently, further improving the construction of single-particle basis sets will significantly help to advance to heavier nuclei and higher accuracies in \emph{ab initio} applications.

\section{Summary and Conclusions}
\label{sec:outlook}

In this work, we performed an extensive study of single-particle bases in nuclear \ai{} applications.
We focused on a set of natural orbitals, Hartree-Fock, and harmonic-oscillator basis states, with natural orbitals based on a perturbatively improved one-body density matrix.
The single-particle wave function and its dependence on the  oscillator frequency as well as the potential occurrence of negative occupations in the NAT basis
were investigated in detail.
Going to sufficiently large model spaces, the natural orbitals provide frequency-independent wave functions for both occupied and unoccupied states.
A reasonable mean-field solution, preferably bound, and a lower-resolution Hamiltonian are key factors to generate a reasonable correlated one-body density matrix and resulting natural orbital basis.
When these conditions are not met, the construction of the natural orbitals does not completely lead to the desired frequency independence and produces states with unphysical negative occupation numbers.
Using the two-body form of the kinetic energy decreased the size of the negative occupations compared to the one- plus two-body form of the kinetic energy.

Hartree-Fock and natural orbital basis states have been benchmarked in medium-mass applications using the IMSRG as a nonperturbative many-body framework.
Comparing ground-state energies and charge radii of medium-mass nuclei, we observed only small differences between the HF and NAT bases for Hamiltonians with only two-body interactions and slightly enhanced variations when three-body forces were included. 
In both cases, the results came closer in agreement by a consistent SRG evolution of \textit{NN} and 3\textit{N} interactions to smaller resolution scales.  
Even though we did not obtain the desired frequency-independent results by applying the NAT basis directly, significantly improved frequency independence and faster model-space convergence were found by constructing the natural orbitals in a large model space and evaluating a subsequent IMSRG evolution in a reduced space.
This strategy presents a promising improvement to advance the reach of \textit{ab initio} methods to heavier nuclei and demonstrates the benefits of investigating the computational basis in more detail.

One possibility for further exploration is the investigation of higher-body contributions and specifically how the difference between the HF and natural orbital basis results arise.
Another direction is to incorporate the natural orbitals in a multireference approach to avoid the single Slater-determinant approximation and fully capitalize on the dynamic correlations included in the perturbatively improved density matrix.

\section*{Acknowledgement}

We thank P.~Arthuis, G.~Hagen, T.~Papenbrock, and R.~Stroberg for useful discussions on this manuscript.
This work is supported in part by the  Deutsche  Forschungsgemeinschaft  (DFG,  German Research Foundation) – Project-ID 279384907 – SFB 1245, and the Max Planck Society.

\bibliography{strongint}

\begin{thebibliography}{68}%
\makeatletter
\providecommand \@ifxundefined [1]{%
 \@ifx{#1\undefined}
}%
\providecommand \@ifnum [1]{%
 \ifnum #1\expandafter \@firstoftwo
 \else \expandafter \@secondoftwo
 \fi
}%
\providecommand \@ifx [1]{%
 \ifx #1\expandafter \@firstoftwo
 \else \expandafter \@secondoftwo
 \fi
}%
\providecommand \natexlab [1]{#1}%
\providecommand \enquote  [1]{``#1''}%
\providecommand \bibnamefont  [1]{#1}%
\providecommand \bibfnamefont [1]{#1}%
\providecommand \citenamefont [1]{#1}%
\providecommand \href@noop [0]{\@secondoftwo}%
\providecommand \href [0]{\begingroup \@sanitize@url \@href}%
\providecommand \@href[1]{\@@startlink{#1}\@@href}%
\providecommand \@@href[1]{\endgroup#1\@@endlink}%
\providecommand \@sanitize@url [0]{\catcode `\\12\catcode `\$12\catcode
  `\&12\catcode `\#12\catcode `\^12\catcode `\_12\catcode `\%12\relax}%
\providecommand \@@startlink[1]{}%
\providecommand \@@endlink[0]{}%
\providecommand \url  [0]{\begingroup\@sanitize@url \@url }%
\providecommand \@url [1]{\endgroup\@href {#1}{\urlprefix }}%
\providecommand \urlprefix  [0]{URL }%
\providecommand \Eprint [0]{\href }%
\providecommand \doibase [0]{http://dx.doi.org/}%
\providecommand \selectlanguage [0]{\@gobble}%
\providecommand \bibinfo  [0]{\@secondoftwo}%
\providecommand \bibfield  [0]{\@secondoftwo}%
\providecommand \translation [1]{[#1]}%
\providecommand \BibitemOpen [0]{}%
\providecommand \bibitemStop [0]{}%
\providecommand \bibitemNoStop [0]{.\EOS\space}%
\providecommand \EOS [0]{\spacefactor3000\relax}%
\providecommand \BibitemShut  [1]{\csname bibitem#1\endcsname}%
\let\auto@bib@innerbib\@empty
\bibitem [{\citenamefont {Hebeler}\ \emph {et~al.}(2015)\citenamefont
  {Hebeler}, \citenamefont {Holt}, \citenamefont {Men{\'e}ndez},\ and\
  \citenamefont {Schwenk}}]{Hebe15ARNPS}%
  \BibitemOpen
  \bibfield  {author} {\bibinfo {author} {\bibfnamefont {K.}~\bibnamefont
  {Hebeler}}, \bibinfo {author} {\bibfnamefont {J.~D.}\ \bibnamefont {Holt}},
  \bibinfo {author} {\bibfnamefont {J.}~\bibnamefont {Men{\'e}ndez}}, \ and\
  \bibinfo {author} {\bibfnamefont {A.}~\bibnamefont {Schwenk}},\ }\bibfield
  {title} {\enquote {\bibinfo {title} {{Nuclear Forces and Their Impact on
  Neutron-Rich Nuclei and Neutron-Rich Matter}},}\ }\href {\doibase
  10.1146/annurev-nucl-102313-025446} {\bibfield  {journal} {\bibinfo
  {journal} {Annu. Rev. Nucl. Part. Sci.}\ }\textbf {\bibinfo {volume} {65}},\
  \bibinfo {pages} {457} (\bibinfo {year} {2015})}\BibitemShut {NoStop}%
\bibitem [{\citenamefont {Hergert}(2020)}]{Herg20review}%
  \BibitemOpen
  \bibfield  {author} {\bibinfo {author} {\bibfnamefont {Heiko}\ \bibnamefont
  {Hergert}},\ }\bibfield  {title} {\enquote {\bibinfo {title} {A {G}uided
  {T}our of ab initio {N}uclear {M}any-{B}ody {T}heory},}\ }\href {\doibase
  10.3389/fphy.2020.00379} {\bibfield  {journal} {\bibinfo  {journal} {Front.
  Phys.}\ }\textbf {\bibinfo {volume} {8}},\ \bibinfo {pages} {379} (\bibinfo
  {year} {2020})}\BibitemShut {NoStop}%
\bibitem [{\citenamefont {Epelbaum}\ \emph {et~al.}(2009)\citenamefont
  {Epelbaum}, \citenamefont {Hammer},\ and\ \citenamefont
  {Mei{\ss}ner}}]{Epel09RMP}%
  \BibitemOpen
  \bibfield  {author} {\bibinfo {author} {\bibfnamefont {E.}~\bibnamefont
  {Epelbaum}}, \bibinfo {author} {\bibfnamefont {H.-W.}\ \bibnamefont
  {Hammer}}, \ and\ \bibinfo {author} {\bibfnamefont {U.-G.}\ \bibnamefont
  {Mei{\ss}ner}},\ }\bibfield  {title} {\enquote {\bibinfo {title} {{Modern
  theory of nuclear forces}},}\ }\href {\doibase 10.1103/RevModPhys.81.1773}
  {\bibfield  {journal} {\bibinfo  {journal} {Rev. Mod. Phys.}\ }\textbf
  {\bibinfo {volume} {81}},\ \bibinfo {pages} {1773} (\bibinfo {year}
  {2009})}\BibitemShut {NoStop}%
\bibitem [{\citenamefont {Machleidt}\ and\ \citenamefont
  {Entem}(2011)}]{Mach11PR}%
  \BibitemOpen
  \bibfield  {author} {\bibinfo {author} {\bibfnamefont {R.}~\bibnamefont
  {Machleidt}}\ and\ \bibinfo {author} {\bibfnamefont {D.~R.}\ \bibnamefont
  {Entem}},\ }\bibfield  {title} {\enquote {\bibinfo {title} {{Chiral effective
  field theory and nuclear forces}},}\ }\href {\doibase
  10.1016/j.physrep.2011.02.001} {\bibfield  {journal} {\bibinfo  {journal}
  {Phys. Rep.}\ }\textbf {\bibinfo {volume} {503}},\ \bibinfo {pages} {1}
  (\bibinfo {year} {2011})}\BibitemShut {NoStop}%
\bibitem [{\citenamefont {Hebeler}\ \emph {et~al.}(2011)\citenamefont
  {Hebeler}, \citenamefont {Bogner}, \citenamefont {Furnstahl}, \citenamefont
  {Nogga},\ and\ \citenamefont {Schwenk}}]{Hebe11fits}%
  \BibitemOpen
  \bibfield  {author} {\bibinfo {author} {\bibfnamefont {K.}~\bibnamefont
  {Hebeler}}, \bibinfo {author} {\bibfnamefont {S.~K.}\ \bibnamefont {Bogner}},
  \bibinfo {author} {\bibfnamefont {R.~J.}\ \bibnamefont {Furnstahl}}, \bibinfo
  {author} {\bibfnamefont {A.}~\bibnamefont {Nogga}}, \ and\ \bibinfo {author}
  {\bibfnamefont {A.}~\bibnamefont {Schwenk}},\ }\bibfield  {title} {\enquote
  {\bibinfo {title} {{Improved nuclear matter calculations from chiral
  low-momentum interactions}},}\ }\href {\doibase 10.1103/PhysRevC.83.031301}
  {\bibfield  {journal} {\bibinfo  {journal} {Phys. Rev. C}\ }\textbf {\bibinfo
  {volume} {83}},\ \bibinfo {pages} {031301(R)} (\bibinfo {year}
  {2011})}\BibitemShut {NoStop}%
\bibitem [{\citenamefont {Ekstr\"om}\ \emph {et~al.}(2015)\citenamefont
  {Ekstr\"om}, \citenamefont {Jansen}, \citenamefont {Wendt}, \citenamefont
  {Hagen}, \citenamefont {Papenbrock}, \citenamefont {Carlsson}, \citenamefont
  {Forss\'en}, \citenamefont {Hjorth-Jensen}, \citenamefont {Navr\'atil},\ and\
  \citenamefont {Nazarewicz}}]{Ekst15sat}%
  \BibitemOpen
  \bibfield  {author} {\bibinfo {author} {\bibfnamefont {A.}~\bibnamefont
  {Ekstr\"om}}, \bibinfo {author} {\bibfnamefont {G.~R.}\ \bibnamefont
  {Jansen}}, \bibinfo {author} {\bibfnamefont {K.~A.}\ \bibnamefont {Wendt}},
  \bibinfo {author} {\bibfnamefont {G.}~\bibnamefont {Hagen}}, \bibinfo
  {author} {\bibfnamefont {T.}~\bibnamefont {Papenbrock}}, \bibinfo {author}
  {\bibfnamefont {B.~D.}\ \bibnamefont {Carlsson}}, \bibinfo {author}
  {\bibfnamefont {C.}~\bibnamefont {Forss\'en}}, \bibinfo {author}
  {\bibfnamefont {M.}~\bibnamefont {Hjorth-Jensen}}, \bibinfo {author}
  {\bibfnamefont {P.}~\bibnamefont {Navr\'atil}}, \ and\ \bibinfo {author}
  {\bibfnamefont {W.}~\bibnamefont {Nazarewicz}},\ }\bibfield  {title}
  {\enquote {\bibinfo {title} {{Accurate nuclear radii and binding energies
  from a chiral interaction}},}\ }\href {\doibase 10.1103/PhysRevC.91.051301}
  {\bibfield  {journal} {\bibinfo  {journal} {Phys. Rev. C}\ }\textbf {\bibinfo
  {volume} {91}},\ \bibinfo {pages} {051301(R)} (\bibinfo {year}
  {2015})}\BibitemShut {NoStop}%
\bibitem [{\citenamefont {Epelbaum}\ \emph {et~al.}(2015)\citenamefont
  {Epelbaum}, \citenamefont {Krebs},\ and\ \citenamefont
  {Mei{\ss}ner}}]{Epel15NNn4lo}%
  \BibitemOpen
  \bibfield  {author} {\bibinfo {author} {\bibfnamefont {E.}~\bibnamefont
  {Epelbaum}}, \bibinfo {author} {\bibfnamefont {H.}~\bibnamefont {Krebs}}, \
  and\ \bibinfo {author} {\bibfnamefont {U.-G.}\ \bibnamefont {Mei{\ss}ner}},\
  }\bibfield  {title} {\enquote {\bibinfo {title} {{Precision Nucleon-Nucleon
  Potential at Fifth Order in the Chiral Expansion}},}\ }\href {\doibase
  10.1103/PhysRevLett.115.122301} {\bibfield  {journal} {\bibinfo  {journal}
  {Phys. Rev. Lett.}\ }\textbf {\bibinfo {volume} {115}},\ \bibinfo {pages}
  {122301} (\bibinfo {year} {2015})}\BibitemShut {NoStop}%
\bibitem [{\citenamefont {Entem}\ \emph {et~al.}(2017)\citenamefont {Entem},
  \citenamefont {Machleidt},\ and\ \citenamefont {Nosyk}}]{Ente17EMn4lo}%
  \BibitemOpen
  \bibfield  {author} {\bibinfo {author} {\bibfnamefont {D.~R.}\ \bibnamefont
  {Entem}}, \bibinfo {author} {\bibfnamefont {R.}~\bibnamefont {Machleidt}}, \
  and\ \bibinfo {author} {\bibfnamefont {Y.}~\bibnamefont {Nosyk}},\ }\bibfield
   {title} {\enquote {\bibinfo {title} {High-quality two-nucleon potentials up
  to fifth order of the chiral expansion},}\ }\href {\doibase
  10.1103/PhysRevC.96.024004} {\bibfield  {journal} {\bibinfo  {journal} {Phys.
  Rev. C}\ }\textbf {\bibinfo {volume} {96}},\ \bibinfo {pages} {024004}
  (\bibinfo {year} {2017})}\BibitemShut {NoStop}%
\bibitem [{\citenamefont {Drischler}\ \emph {et~al.}(2019)\citenamefont
  {Drischler}, \citenamefont {Hebeler},\ and\ \citenamefont
  {Schwenk}}]{Dris17MCshort}%
  \BibitemOpen
  \bibfield  {author} {\bibinfo {author} {\bibfnamefont {C.}~\bibnamefont
  {Drischler}}, \bibinfo {author} {\bibfnamefont {K.}~\bibnamefont {Hebeler}},
  \ and\ \bibinfo {author} {\bibfnamefont {A.}~\bibnamefont {Schwenk}},\
  }\bibfield  {title} {\enquote {\bibinfo {title} {{Chiral {I}nteractions up to
  {N}ext-to-{N}ext-to-{N}ext-to-{L}eading {O}rder and {N}uclear
  {S}aturation}},}\ }\href {\doibase 10.1103/PhysRevLett.122.042501} {\bibfield
   {journal} {\bibinfo  {journal} {Phys. Rev. Lett.}\ }\textbf {\bibinfo
  {volume} {122}},\ \bibinfo {pages} {042501} (\bibinfo {year}
  {2019})}\BibitemShut {NoStop}%
\bibitem [{\citenamefont {Hoppe}\ \emph {et~al.}(2019)\citenamefont {Hoppe},
  \citenamefont {Drischler}, \citenamefont {Hebeler}, \citenamefont {Schwenk},\
  and\ \citenamefont {Simonis}}]{Hopp19medmass}%
  \BibitemOpen
  \bibfield  {author} {\bibinfo {author} {\bibfnamefont {J.}~\bibnamefont
  {Hoppe}}, \bibinfo {author} {\bibfnamefont {C.}~\bibnamefont {Drischler}},
  \bibinfo {author} {\bibfnamefont {K.}~\bibnamefont {Hebeler}}, \bibinfo
  {author} {\bibfnamefont {A.}~\bibnamefont {Schwenk}}, \ and\ \bibinfo
  {author} {\bibfnamefont {J.}~\bibnamefont {Simonis}},\ }\bibfield  {title}
  {\enquote {\bibinfo {title} {{Probing chiral interactions up to
  next-to-next-to-next-to-leading order in medium-mass nuclei}},}\ }\href
  {\doibase 10.1103/PhysRevC.100.024318} {\bibfield  {journal} {\bibinfo
  {journal} {Phys. Rev. C}\ }\textbf {\bibinfo {volume} {100}},\ \bibinfo
  {pages} {024318} (\bibinfo {year} {2019})}\BibitemShut {NoStop}%
\bibitem [{\citenamefont {Hüther}\ \emph {et~al.}(2020)\citenamefont
  {Hüther}, \citenamefont {Vobig}, \citenamefont {Hebeler}, \citenamefont
  {Machleidt},\ and\ \citenamefont {Roth}}]{Huth19chiralfam}%
  \BibitemOpen
  \bibfield  {author} {\bibinfo {author} {\bibfnamefont {T.}~\bibnamefont
  {Hüther}}, \bibinfo {author} {\bibfnamefont {K.}~\bibnamefont {Vobig}},
  \bibinfo {author} {\bibfnamefont {K.}~\bibnamefont {Hebeler}}, \bibinfo
  {author} {\bibfnamefont {R.}~\bibnamefont {Machleidt}}, \ and\ \bibinfo
  {author} {\bibfnamefont {R.}~\bibnamefont {Roth}},\ }\bibfield  {title}
  {\enquote {\bibinfo {title} {Family of chiral two- plus three-nucleon
  interactions for accurate nuclear structure studies},}\ }\href {\doibase
  https://doi.org/10.1016/j.physletb.2020.135651} {\bibfield  {journal}
  {\bibinfo  {journal} {Phys. Lett. B}\ }\textbf {\bibinfo {volume} {808}},\
  \bibinfo {pages} {135651} (\bibinfo {year} {2020})}\BibitemShut {NoStop}%
\bibitem [{\citenamefont {Epelbaum}\ \emph {et~al.}(2020)\citenamefont
  {Epelbaum}, \citenamefont {Krebs},\ and\ \citenamefont
  {Reinert}}]{Epel19nuclfFront}%
  \BibitemOpen
  \bibfield  {author} {\bibinfo {author} {\bibfnamefont {E.}~\bibnamefont
  {Epelbaum}}, \bibinfo {author} {\bibfnamefont {H.}~\bibnamefont {Krebs}}, \
  and\ \bibinfo {author} {\bibfnamefont {P}~\bibnamefont {Reinert}},\
  }\bibfield  {title} {\enquote {\bibinfo {title} {{High-Precision Nuclear
  Forces From Chiral EFT: State-of-the-Art, Challenges, and Outlook}},}\ }\href
  {\doibase 10.3389/fphy.2020.00098} {\bibfield  {journal} {\bibinfo  {journal}
  {Front. Phys.}\ }\textbf {\bibinfo {volume} {8}},\ \bibinfo {pages} {98}
  (\bibinfo {year} {2020})}\BibitemShut {NoStop}%
\bibitem [{\citenamefont {Hebeler}(2021)}]{Hebe203NF}%
  \BibitemOpen
  \bibfield  {author} {\bibinfo {author} {\bibfnamefont {Kai}\ \bibnamefont
  {Hebeler}},\ }\bibfield  {title} {\enquote {\bibinfo {title} {Three-nucleon
  forces: Implementation and applications to atomic nuclei and dense matter},}\
  }\href {\doibase https://doi.org/10.1016/j.physrep.2020.08.009} {\bibfield
  {journal} {\bibinfo  {journal} {Phys. Rep.}\ }\textbf {\bibinfo {volume}
  {890}},\ \bibinfo {pages} {1} (\bibinfo {year} {2021})}\BibitemShut {NoStop}%
\bibitem [{\citenamefont {Jiang}\ \emph {et~al.}(2020)\citenamefont {Jiang},
  \citenamefont {Ekstr\"om}, \citenamefont {Forss\'en}, \citenamefont {Hagen},
  \citenamefont {Jansen},\ and\ \citenamefont {Papenbrock}}]{Jian20N2LOGO}%
  \BibitemOpen
  \bibfield  {author} {\bibinfo {author} {\bibfnamefont {W.~G.}\ \bibnamefont
  {Jiang}}, \bibinfo {author} {\bibfnamefont {A.}~\bibnamefont {Ekstr\"om}},
  \bibinfo {author} {\bibfnamefont {C.}~\bibnamefont {Forss\'en}}, \bibinfo
  {author} {\bibfnamefont {G.}~\bibnamefont {Hagen}}, \bibinfo {author}
  {\bibfnamefont {G.~R.}\ \bibnamefont {Jansen}}, \ and\ \bibinfo {author}
  {\bibfnamefont {T.}~\bibnamefont {Papenbrock}},\ }\bibfield  {title}
  {\enquote {\bibinfo {title} {Accurate bulk properties of nuclei from ${A}=2$
  to $\ensuremath{\infty}$ from potentials with $\mathrm{\ensuremath{\Delta}}$
  isobars},}\ }\href {\doibase 10.1103/PhysRevC.102.054301} {\bibfield
  {journal} {\bibinfo  {journal} {Phys. Rev. C}\ }\textbf {\bibinfo {volume}
  {102}},\ \bibinfo {pages} {054301} (\bibinfo {year} {2020})}\BibitemShut
  {NoStop}%
\bibitem [{\citenamefont {Dickhoff}\ and\ \citenamefont
  {Barbieri}(2004)}]{Dick04PPNP}%
  \BibitemOpen
  \bibfield  {author} {\bibinfo {author} {\bibfnamefont {W.~H.}\ \bibnamefont
  {Dickhoff}}\ and\ \bibinfo {author} {\bibfnamefont {C.}~\bibnamefont
  {Barbieri}},\ }\bibfield  {title} {\enquote {\bibinfo {title}
  {{Self-consistent Green's function method for nuclei and nuclear matter}},}\
  }\href {\doibase 10.1016/j.ppnp.2004.02.038} {\bibfield  {journal} {\bibinfo
  {journal} {Prog. Part. Nucl. Phys.}\ }\textbf {\bibinfo {volume} {52}},\
  \bibinfo {pages} {377} (\bibinfo {year} {2004})}\BibitemShut {NoStop}%
\bibitem [{\citenamefont {Hagen}\ \emph {et~al.}(2014)\citenamefont {Hagen},
  \citenamefont {Papenbrock}, \citenamefont {Hjorth-Jensen},\ and\
  \citenamefont {Dean}}]{Hage14RPP}%
  \BibitemOpen
  \bibfield  {author} {\bibinfo {author} {\bibfnamefont {G.}~\bibnamefont
  {Hagen}}, \bibinfo {author} {\bibfnamefont {T.}~\bibnamefont {Papenbrock}},
  \bibinfo {author} {\bibfnamefont {M.}~\bibnamefont {Hjorth-Jensen}}, \ and\
  \bibinfo {author} {\bibfnamefont {D.~J.}\ \bibnamefont {Dean}},\ }\bibfield
  {title} {\enquote {\bibinfo {title} {{Coupled-cluster computations of atomic
  nuclei}},}\ }\href {\doibase 10.1088/0034-4885/77/9/096302} {\bibfield
  {journal} {\bibinfo  {journal} {Rep. Prog. Phys.}\ }\textbf {\bibinfo
  {volume} {77}},\ \bibinfo {pages} {096302} (\bibinfo {year}
  {2014})}\BibitemShut {NoStop}%
\bibitem [{\citenamefont {Hergert}\ \emph {et~al.}(2016)\citenamefont
  {Hergert}, \citenamefont {Bogner}, \citenamefont {Morris}, \citenamefont
  {Schwenk},\ and\ \citenamefont {Tsukiyama}}]{Herg16PR}%
  \BibitemOpen
  \bibfield  {author} {\bibinfo {author} {\bibfnamefont {H.}~\bibnamefont
  {Hergert}}, \bibinfo {author} {\bibfnamefont {S.~K.}\ \bibnamefont {Bogner}},
  \bibinfo {author} {\bibfnamefont {T.~D.}\ \bibnamefont {Morris}}, \bibinfo
  {author} {\bibfnamefont {A.}~\bibnamefont {Schwenk}}, \ and\ \bibinfo
  {author} {\bibfnamefont {K.}~\bibnamefont {Tsukiyama}},\ }\bibfield  {title}
  {\enquote {\bibinfo {title} {{The In-Medium Similarity Renormalization Group:
  A Novel Ab Initio Method for Nuclei}},}\ }\href {\doibase
  10.1016/j.physrep.2015.12.007} {\bibfield  {journal} {\bibinfo  {journal}
  {Phys. Rept.}\ }\textbf {\bibinfo {volume} {621}},\ \bibinfo {pages} {165}
  (\bibinfo {year} {2016})}\BibitemShut {NoStop}%
\bibitem [{\citenamefont {Tichai}\ \emph {et~al.}(2020)\citenamefont {Tichai},
  \citenamefont {Roth},\ and\ \citenamefont {Duguet}}]{Tichai2020review}%
  \BibitemOpen
  \bibfield  {author} {\bibinfo {author} {\bibfnamefont {A.}~\bibnamefont
  {Tichai}}, \bibinfo {author} {\bibfnamefont {R.}~\bibnamefont {Roth}}, \ and\
  \bibinfo {author} {\bibfnamefont {T.}~\bibnamefont {Duguet}},\ }\bibfield
  {title} {\enquote {\bibinfo {title} {Many-{B}ody {P}erturbation {T}heories
  for {F}inite {N}uclei},}\ }\href {\doibase 10.3389/fphy.2020.00164}
  {\bibfield  {journal} {\bibinfo  {journal} {Front. Phys.}\ }\textbf {\bibinfo
  {volume} {8}},\ \bibinfo {pages} {164} (\bibinfo {year} {2020})}\BibitemShut
  {NoStop}%
\bibitem [{\citenamefont {Morris}\ \emph {et~al.}(2018)\citenamefont {Morris},
  \citenamefont {Simonis}, \citenamefont {Stroberg}, \citenamefont {Stumpf},
  \citenamefont {Hagen}, \citenamefont {Holt}, \citenamefont {Jansen},
  \citenamefont {Papenbrock}, \citenamefont {Roth},\ and\ \citenamefont
  {Schwenk}}]{Morr17Tin}%
  \BibitemOpen
  \bibfield  {author} {\bibinfo {author} {\bibfnamefont {T.~D.}\ \bibnamefont
  {Morris}}, \bibinfo {author} {\bibfnamefont {J.}~\bibnamefont {Simonis}},
  \bibinfo {author} {\bibfnamefont {S.~R.}\ \bibnamefont {Stroberg}}, \bibinfo
  {author} {\bibfnamefont {C.}~\bibnamefont {Stumpf}}, \bibinfo {author}
  {\bibfnamefont {G.}~\bibnamefont {Hagen}}, \bibinfo {author} {\bibfnamefont
  {J.~D.}\ \bibnamefont {Holt}}, \bibinfo {author} {\bibfnamefont {G.~R.}\
  \bibnamefont {Jansen}}, \bibinfo {author} {\bibfnamefont {T.}~\bibnamefont
  {Papenbrock}}, \bibinfo {author} {\bibfnamefont {R.}~\bibnamefont {Roth}}, \
  and\ \bibinfo {author} {\bibfnamefont {A.}~\bibnamefont {Schwenk}},\
  }\bibfield  {title} {\enquote {\bibinfo {title} {Structure of the {L}ightest
  {T}in {I}sotopes},}\ }\href {\doibase 10.1103/PhysRevLett.120.152503}
  {\bibfield  {journal} {\bibinfo  {journal} {Phys. Rev. Lett.}\ }\textbf
  {\bibinfo {volume} {120}},\ \bibinfo {pages} {152503} (\bibinfo {year}
  {2018})}\BibitemShut {NoStop}%
\bibitem [{\citenamefont {Holt}\ \emph {et~al.}(2014)\citenamefont {Holt},
  \citenamefont {Men{\'e}ndez}, \citenamefont {Simonis},\ and\ \citenamefont
  {Schwenk}}]{Holt14Ca}%
  \BibitemOpen
  \bibfield  {author} {\bibinfo {author} {\bibfnamefont {J.~D.}\ \bibnamefont
  {Holt}}, \bibinfo {author} {\bibfnamefont {J.}~\bibnamefont {Men{\'e}ndez}},
  \bibinfo {author} {\bibfnamefont {J.}~\bibnamefont {Simonis}}, \ and\
  \bibinfo {author} {\bibfnamefont {A.}~\bibnamefont {Schwenk}},\ }\bibfield
  {title} {\enquote {\bibinfo {title} {{Three-nucleon forces and spectroscopy
  of neutron-rich calcium isotopes}},}\ }\href {\doibase
  10.1103/PhysRevC.90.024312} {\bibfield  {journal} {\bibinfo  {journal} {Phys.
  Rev. C}\ }\textbf {\bibinfo {volume} {90}},\ \bibinfo {pages} {024312}
  (\bibinfo {year} {2014})}\BibitemShut {NoStop}%
\bibitem [{\citenamefont {Tichai}\ \emph {et~al.}(2016)\citenamefont {Tichai},
  \citenamefont {Langhammer}, \citenamefont {Binder},\ and\ \citenamefont
  {Roth}}]{Tich16HFMBPT}%
  \BibitemOpen
  \bibfield  {author} {\bibinfo {author} {\bibfnamefont {A.}~\bibnamefont
  {Tichai}}, \bibinfo {author} {\bibfnamefont {J.}~\bibnamefont {Langhammer}},
  \bibinfo {author} {\bibfnamefont {S.}~\bibnamefont {Binder}}, \ and\ \bibinfo
  {author} {\bibfnamefont {R.}~\bibnamefont {Roth}},\ }\bibfield  {title}
  {\enquote {\bibinfo {title} {{Hartree-Fock many-body perturbation theory for
  nuclear ground-states}},}\ }\href {\doibase 10.1016/j.physletb.2016.03.029}
  {\bibfield  {journal} {\bibinfo  {journal} {Phys. Lett. B}\ }\textbf
  {\bibinfo {volume} {756}},\ \bibinfo {pages} {283} (\bibinfo {year}
  {2016})}\BibitemShut {NoStop}%
\bibitem [{\citenamefont {Tichai}\ \emph {et~al.}(2018)\citenamefont {Tichai},
  \citenamefont {Arthuis}, \citenamefont {Duguet}, \citenamefont {Hergert},
  \citenamefont {Somá},\ and\ \citenamefont {Roth}}]{Tichai18BMBPT}%
  \BibitemOpen
  \bibfield  {author} {\bibinfo {author} {\bibfnamefont {A.}~\bibnamefont
  {Tichai}}, \bibinfo {author} {\bibfnamefont {P.}~\bibnamefont {Arthuis}},
  \bibinfo {author} {\bibfnamefont {T.}~\bibnamefont {Duguet}}, \bibinfo
  {author} {\bibfnamefont {H.}~\bibnamefont {Hergert}}, \bibinfo {author}
  {\bibfnamefont {V.}~\bibnamefont {Somá}}, \ and\ \bibinfo {author}
  {\bibfnamefont {R.}~\bibnamefont {Roth}},\ }\bibfield  {title} {\enquote
  {\bibinfo {title} {{Bogoliubov many-body perturbation theory for open-shell
  nuclei}},}\ }\href {\doibase 10.1016/j.physletb.2018.09.044} {\bibfield
  {journal} {\bibinfo  {journal} {Phys. Lett. B}\ }\textbf {\bibinfo {volume}
  {786}},\ \bibinfo {pages} {195} (\bibinfo {year} {2018})}\BibitemShut
  {NoStop}%
\bibitem [{\citenamefont {Binder}\ \emph {et~al.}(2014)\citenamefont {Binder},
  \citenamefont {Langhammer}, \citenamefont {Calci},\ and\ \citenamefont
  {Roth}}]{Bind14CCheavy}%
  \BibitemOpen
  \bibfield  {author} {\bibinfo {author} {\bibfnamefont {S.}~\bibnamefont
  {Binder}}, \bibinfo {author} {\bibfnamefont {J.}~\bibnamefont {Langhammer}},
  \bibinfo {author} {\bibfnamefont {A.}~\bibnamefont {Calci}}, \ and\ \bibinfo
  {author} {\bibfnamefont {R.}~\bibnamefont {Roth}},\ }\bibfield  {title}
  {\enquote {\bibinfo {title} {{Ab initio path to heavy nuclei}},}\ }\href
  {\doibase 10.1016/j.physletb.2014.07.010} {\bibfield  {journal} {\bibinfo
  {journal} {Phys. Lett. B}\ }\textbf {\bibinfo {volume} {736}},\ \bibinfo
  {pages} {119} (\bibinfo {year} {2014})}\BibitemShut {NoStop}%
\bibitem [{\citenamefont {Tsukiyama}\ \emph {et~al.}(2011)\citenamefont
  {Tsukiyama}, \citenamefont {Bogner},\ and\ \citenamefont
  {Schwenk}}]{Tsuk11IMSRG}%
  \BibitemOpen
  \bibfield  {author} {\bibinfo {author} {\bibfnamefont {K.}~\bibnamefont
  {Tsukiyama}}, \bibinfo {author} {\bibfnamefont {S.~K.}\ \bibnamefont
  {Bogner}}, \ and\ \bibinfo {author} {\bibfnamefont {A.}~\bibnamefont
  {Schwenk}},\ }\bibfield  {title} {\enquote {\bibinfo {title} {{In-medium
  Similarity Renormalization Group for Nuclei}},}\ }\href {\doibase
  10.1103/PhysRevLett.106.222502} {\bibfield  {journal} {\bibinfo  {journal}
  {Phys. Rev. Lett.}\ }\textbf {\bibinfo {volume} {106}},\ \bibinfo {pages}
  {222502} (\bibinfo {year} {2011})}\BibitemShut {NoStop}%
\bibitem [{\citenamefont {Stroberg}\ \emph {et~al.}(2019)\citenamefont
  {Stroberg}, \citenamefont {Hergert}, \citenamefont {Bogner},\ and\
  \citenamefont {Holt}}]{Stroberg2019}%
  \BibitemOpen
  \bibfield  {author} {\bibinfo {author} {\bibfnamefont {S.~R.}\ \bibnamefont
  {Stroberg}}, \bibinfo {author} {\bibfnamefont {H.}~\bibnamefont {Hergert}},
  \bibinfo {author} {\bibfnamefont {S.~K.}\ \bibnamefont {Bogner}}, \ and\
  \bibinfo {author} {\bibfnamefont {J.~D.}\ \bibnamefont {Holt}},\ }\bibfield
  {title} {\enquote {\bibinfo {title} {Nonempirical {I}nteractions for the
  {N}uclear {S}hell {M}odel: An {U}pdate},}\ }\href {\doibase
  10.1146/annurev-nucl-101917-021120} {\bibfield  {journal} {\bibinfo
  {journal} {Annu. Rev. Nucl. Part. Sci.}\ }\textbf {\bibinfo {volume} {69}},\
  \bibinfo {pages} {307} (\bibinfo {year} {2019})}\BibitemShut {NoStop}%
\bibitem [{\citenamefont {Som{\`a}}\ \emph {et~al.}(2020)\citenamefont
  {Som{\`a}}, \citenamefont {Navr{\'a}til}, \citenamefont {Raimondi},
  \citenamefont {Barbieri},\ and\ \citenamefont {Duguet}}]{Soma20SCGF}%
  \BibitemOpen
  \bibfield  {author} {\bibinfo {author} {\bibfnamefont {V.}~\bibnamefont
  {Som{\`a}}}, \bibinfo {author} {\bibfnamefont {P.}~\bibnamefont
  {Navr{\'a}til}}, \bibinfo {author} {\bibfnamefont {F.}~\bibnamefont
  {Raimondi}}, \bibinfo {author} {\bibfnamefont {C.}~\bibnamefont {Barbieri}},
  \ and\ \bibinfo {author} {\bibfnamefont {T.}~\bibnamefont {Duguet}},\
  }\bibfield  {title} {\enquote {\bibinfo {title} {{Novel chiral Hamiltonian
  and observables in light and medium-mass nuclei}},}\ }\href {\doibase
  10.1103/PhysRevC.101.014318} {\bibfield  {journal} {\bibinfo  {journal}
  {Phys. Rev. C}\ }\textbf {\bibinfo {volume} {101}},\ \bibinfo {pages}
  {014318} (\bibinfo {year} {2020})}\BibitemShut {NoStop}%
\bibitem [{\citenamefont {L{\"a}hde}\ \emph {et~al.}(2014)\citenamefont
  {L{\"a}hde}, \citenamefont {Epelbaum}, \citenamefont {Krebs}, \citenamefont
  {Lee}, \citenamefont {Mei{\ss}ner},\ and\ \citenamefont
  {Rupak}}]{Lahd13LEFT}%
  \BibitemOpen
  \bibfield  {author} {\bibinfo {author} {\bibfnamefont {T.~A.}\ \bibnamefont
  {L{\"a}hde}}, \bibinfo {author} {\bibfnamefont {E.}~\bibnamefont {Epelbaum}},
  \bibinfo {author} {\bibfnamefont {H.}~\bibnamefont {Krebs}}, \bibinfo
  {author} {\bibfnamefont {D.}~\bibnamefont {Lee}}, \bibinfo {author}
  {\bibfnamefont {U.-G.}\ \bibnamefont {Mei{\ss}ner}}, \ and\ \bibinfo {author}
  {\bibfnamefont {G.}~\bibnamefont {Rupak}},\ }\bibfield  {title} {\enquote
  {\bibinfo {title} {{Lattice effective field theory for medium-mass
  nuclei}},}\ }\href {\doibase 10.1016/j.physletb.2014.03.023} {\bibfield
  {journal} {\bibinfo  {journal} {Phys. Lett. B}\ }\textbf {\bibinfo {volume}
  {732}},\ \bibinfo {pages} {110} (\bibinfo {year} {2014})}\BibitemShut
  {NoStop}%
\bibitem [{\citenamefont {Hagen}\ \emph {et~al.}(2016)\citenamefont {Hagen},
  \citenamefont {Ekstr{\"o}m}, \citenamefont {Forss{\'e}n}, \citenamefont
  {Jansen}, \citenamefont {Nazarewicz}, \citenamefont {Papenbrock},
  \citenamefont {Wendt}, \citenamefont {Bacca}, \citenamefont {Barnea},
  \citenamefont {Carlsson}, \citenamefont {Drischler}, \citenamefont {Hebeler},
  \citenamefont {Hjorth-Jensen}, \citenamefont {Miorelli}, \citenamefont
  {Orlandini}, \citenamefont {Schwenk},\ and\ \citenamefont
  {Simonis}}]{Hage16NatPhys}%
  \BibitemOpen
  \bibfield  {author} {\bibinfo {author} {\bibfnamefont {G.}~\bibnamefont
  {Hagen}}, \bibinfo {author} {\bibfnamefont {A.}~\bibnamefont {Ekstr{\"o}m}},
  \bibinfo {author} {\bibfnamefont {C.}~\bibnamefont {Forss{\'e}n}}, \bibinfo
  {author} {\bibfnamefont {G.~R.}\ \bibnamefont {Jansen}}, \bibinfo {author}
  {\bibfnamefont {W.}~\bibnamefont {Nazarewicz}}, \bibinfo {author}
  {\bibfnamefont {T.}~\bibnamefont {Papenbrock}}, \bibinfo {author}
  {\bibfnamefont {K.~A.}\ \bibnamefont {Wendt}}, \bibinfo {author}
  {\bibfnamefont {S.}~\bibnamefont {Bacca}}, \bibinfo {author} {\bibfnamefont
  {N.}~\bibnamefont {Barnea}}, \bibinfo {author} {\bibfnamefont
  {B.}~\bibnamefont {Carlsson}}, \bibinfo {author} {\bibfnamefont
  {C.}~\bibnamefont {Drischler}}, \bibinfo {author} {\bibfnamefont
  {K.}~\bibnamefont {Hebeler}}, \bibinfo {author} {\bibfnamefont
  {M.}~\bibnamefont {Hjorth-Jensen}}, \bibinfo {author} {\bibfnamefont
  {M.}~\bibnamefont {Miorelli}}, \bibinfo {author} {\bibfnamefont
  {G.}~\bibnamefont {Orlandini}}, \bibinfo {author} {\bibfnamefont
  {A.}~\bibnamefont {Schwenk}}, \ and\ \bibinfo {author} {\bibfnamefont
  {J.}~\bibnamefont {Simonis}},\ }\bibfield  {title} {\enquote {\bibinfo
  {title} {{Neutron and weak-charge distributions of the $^{48}$Ca nucleus}},}\
  }\href {\doibase 10.1038/nphys3529} {\bibfield  {journal} {\bibinfo
  {journal} {Nat. Phys.}\ }\textbf {\bibinfo {volume} {12}},\ \bibinfo {pages}
  {186} (\bibinfo {year} {2016})}\BibitemShut {NoStop}%
\bibitem [{\citenamefont {Gysbers}\ \emph {et~al.}(2019)\citenamefont
  {Gysbers}, \citenamefont {Hagen}, \citenamefont {Holt}, \citenamefont
  {Jansen}, \citenamefont {Morris}, \citenamefont {Navr{\'a}til}, \citenamefont
  {Papenbrock}, \citenamefont {Quaglioni}, \citenamefont {Schwenk},
  \citenamefont {Stroberg},\ and\ \citenamefont {Wendt}}]{Gysb19beta}%
  \BibitemOpen
  \bibfield  {author} {\bibinfo {author} {\bibfnamefont {P.}~\bibnamefont
  {Gysbers}}, \bibinfo {author} {\bibfnamefont {G.}~\bibnamefont {Hagen}},
  \bibinfo {author} {\bibfnamefont {J.~D.}\ \bibnamefont {Holt}}, \bibinfo
  {author} {\bibfnamefont {G.~R.}\ \bibnamefont {Jansen}}, \bibinfo {author}
  {\bibfnamefont {T.~D.}\ \bibnamefont {Morris}}, \bibinfo {author}
  {\bibfnamefont {P.}~\bibnamefont {Navr{\'a}til}}, \bibinfo {author}
  {\bibfnamefont {T.}~\bibnamefont {Papenbrock}}, \bibinfo {author}
  {\bibfnamefont {S.}~\bibnamefont {Quaglioni}}, \bibinfo {author}
  {\bibfnamefont {A.}~\bibnamefont {Schwenk}}, \bibinfo {author} {\bibfnamefont
  {S.~R.}\ \bibnamefont {Stroberg}}, \ and\ \bibinfo {author} {\bibfnamefont
  {K.~A.}\ \bibnamefont {Wendt}},\ }\bibfield  {title} {\enquote {\bibinfo
  {title} {{Discrepancy between experimental and theoretical $\beta$-decay
  rates resolved from first principles}},}\ }\href {\doibase
  10.1038/s41567-019-0450-7} {\bibfield  {journal} {\bibinfo  {journal} {Nat.
  Phys.}\ }\textbf {\bibinfo {volume} {15}},\ \bibinfo {pages} {428} (\bibinfo
  {year} {2019})}\BibitemShut {NoStop}%
\bibitem [{\citenamefont {Yao}\ \emph {et~al.}(2020)\citenamefont {Yao},
  \citenamefont {Bally}, \citenamefont {Engel}, \citenamefont {Wirth},
  \citenamefont {Rodr\'{\i}guez},\ and\ \citenamefont {Hergert}}]{Yao2020}%
  \BibitemOpen
  \bibfield  {author} {\bibinfo {author} {\bibfnamefont {J.~M.}\ \bibnamefont
  {Yao}}, \bibinfo {author} {\bibfnamefont {B.}~\bibnamefont {Bally}}, \bibinfo
  {author} {\bibfnamefont {J.}~\bibnamefont {Engel}}, \bibinfo {author}
  {\bibfnamefont {R.}~\bibnamefont {Wirth}}, \bibinfo {author} {\bibfnamefont
  {T.~R.}\ \bibnamefont {Rodr\'{\i}guez}}, \ and\ \bibinfo {author}
  {\bibfnamefont {H.}~\bibnamefont {Hergert}},\ }\bibfield  {title} {\enquote
  {\bibinfo {title} {Ab {I}nitio {T}reatment of {C}ollective {C}orrelations and
  the {N}eutrinoless {D}ouble {B}eta {D}ecay of $^{48}\mathrm{Ca}$},}\ }\href
  {\doibase 10.1103/PhysRevLett.124.232501} {\bibfield  {journal} {\bibinfo
  {journal} {Phys. Rev. Lett.}\ }\textbf {\bibinfo {volume} {124}},\ \bibinfo
  {pages} {232501} (\bibinfo {year} {2020})}\BibitemShut {NoStop}%
\bibitem [{\citenamefont {Caprio}\ \emph {et~al.}(2012)\citenamefont {Caprio},
  \citenamefont {Maris},\ and\ \citenamefont {Vary}}]{Capr12Sturmian}%
  \BibitemOpen
  \bibfield  {author} {\bibinfo {author} {\bibfnamefont {M.~A.}\ \bibnamefont
  {Caprio}}, \bibinfo {author} {\bibfnamefont {P.}~\bibnamefont {Maris}}, \
  and\ \bibinfo {author} {\bibfnamefont {J.~P.}\ \bibnamefont {Vary}},\
  }\bibfield  {title} {\enquote {\bibinfo {title} {Coulomb-sturmian basis for
  the nuclear many-body problem},}\ }\href {\doibase
  10.1103/PhysRevC.86.034312} {\bibfield  {journal} {\bibinfo  {journal} {Phys.
  Rev. C}\ }\textbf {\bibinfo {volume} {86}},\ \bibinfo {pages} {034312}
  (\bibinfo {year} {2012})}\BibitemShut {NoStop}%
\bibitem [{\citenamefont {Tichai}\ \emph {et~al.}(2019)\citenamefont {Tichai},
  \citenamefont {M{\"u}ller}, \citenamefont {Vobig},\ and\ \citenamefont
  {Roth}}]{Tich19NatNCSM}%
  \BibitemOpen
  \bibfield  {author} {\bibinfo {author} {\bibfnamefont {A.}~\bibnamefont
  {Tichai}}, \bibinfo {author} {\bibfnamefont {J.}~\bibnamefont {M{\"u}ller}},
  \bibinfo {author} {\bibfnamefont {K.}~\bibnamefont {Vobig}}, \ and\ \bibinfo
  {author} {\bibfnamefont {R.}~\bibnamefont {Roth}},\ }\bibfield  {title}
  {\enquote {\bibinfo {title} {{Natural orbitals for ab initio no-core shell
  model calculations}},}\ }\href {\doibase 10.1103/PhysRevC.99.034321}
  {\bibfield  {journal} {\bibinfo  {journal} {Phys. Rev. C}\ }\textbf {\bibinfo
  {volume} {99}},\ \bibinfo {pages} {034321} (\bibinfo {year}
  {2019})}\BibitemShut {NoStop}%
\bibitem [{\citenamefont {Navratil}\ \emph {et~al.}(2009)\citenamefont
  {Navratil}, \citenamefont {Quaglioni}, \citenamefont {Stetcu},\ and\
  \citenamefont {Barrett}}]{Navr09NCSMdev}%
  \BibitemOpen
  \bibfield  {author} {\bibinfo {author} {\bibfnamefont {P.}~\bibnamefont
  {Navratil}}, \bibinfo {author} {\bibfnamefont {S.}~\bibnamefont {Quaglioni}},
  \bibinfo {author} {\bibfnamefont {I.}~\bibnamefont {Stetcu}}, \ and\ \bibinfo
  {author} {\bibfnamefont {B.~R.}\ \bibnamefont {Barrett}},\ }\bibfield
  {title} {\enquote {\bibinfo {title} {{Recent developments in no-core
  shell-model calculations}},}\ }\href {\doibase 10.1088/0954-3899/36/8/083101}
  {\bibfield  {journal} {\bibinfo  {journal} {J. Phys. G}\ }\textbf {\bibinfo
  {volume} {36}},\ \bibinfo {pages} {083101} (\bibinfo {year}
  {2009})}\BibitemShut {NoStop}%
\bibitem [{\citenamefont {Barrett}\ \emph {et~al.}(2013)\citenamefont
  {Barrett}, \citenamefont {Navr{\'a}til},\ and\ \citenamefont
  {Vary}}]{Barr13PPNP}%
  \BibitemOpen
  \bibfield  {author} {\bibinfo {author} {\bibfnamefont {B.~R.}\ \bibnamefont
  {Barrett}}, \bibinfo {author} {\bibfnamefont {P.}~\bibnamefont
  {Navr{\'a}til}}, \ and\ \bibinfo {author} {\bibfnamefont {J.~P.}\
  \bibnamefont {Vary}},\ }\bibfield  {title} {\enquote {\bibinfo {title} {{Ab
  initio no core shell model}},}\ }\href {\doibase 10.1016/j.ppnp.2012.10.003}
  {\bibfield  {journal} {\bibinfo  {journal} {Prog. Part. Nucl. Phys.}\
  }\textbf {\bibinfo {volume} {69}},\ \bibinfo {pages} {131} (\bibinfo {year}
  {2013})}\BibitemShut {NoStop}%
\bibitem [{\citenamefont {Hagen}\ \emph {et~al.}(2009)\citenamefont {Hagen},
  \citenamefont {Papenbrock},\ and\ \citenamefont {Dean}}]{Hage09CoM}%
  \BibitemOpen
  \bibfield  {author} {\bibinfo {author} {\bibfnamefont {G.}~\bibnamefont
  {Hagen}}, \bibinfo {author} {\bibfnamefont {T.}~\bibnamefont {Papenbrock}}, \
  and\ \bibinfo {author} {\bibfnamefont {D.~J.}\ \bibnamefont {Dean}},\
  }\bibfield  {title} {\enquote {\bibinfo {title} {{Solution of the
  {C}enter-{O}f-{M}ass {P}roblem in {N}uclear {S}tructure {C}alculations}},}\
  }\href {\doibase 10.1103/PhysRevLett.103.062503} {\bibfield  {journal}
  {\bibinfo  {journal} {Phys. Rev. Lett.}\ }\textbf {\bibinfo {volume} {103}},\
  \bibinfo {pages} {062503} (\bibinfo {year} {2009})}\BibitemShut {NoStop}%
\bibitem [{\citenamefont {Novario}\ \emph
  {et~al.}(2020{\natexlab{a}})\citenamefont {Novario}, \citenamefont {Hagen},
  \citenamefont {Jansen},\ and\ \citenamefont {Papenbrock}}]{Novario2020a}%
  \BibitemOpen
  \bibfield  {author} {\bibinfo {author} {\bibfnamefont {S.~J.}\ \bibnamefont
  {Novario}}, \bibinfo {author} {\bibfnamefont {G.}~\bibnamefont {Hagen}},
  \bibinfo {author} {\bibfnamefont {G.~R.}\ \bibnamefont {Jansen}}, \ and\
  \bibinfo {author} {\bibfnamefont {T.}~\bibnamefont {Papenbrock}},\ }\bibfield
   {title} {\enquote {\bibinfo {title} {Charge radii of exotic neon and
  magnesium isotopes},}\ }\href {\doibase 10.1103/PhysRevC.102.051303}
  {\bibfield  {journal} {\bibinfo  {journal} {Phys. Rev. C}\ }\textbf {\bibinfo
  {volume} {102}},\ \bibinfo {pages} {051303(R)} (\bibinfo {year}
  {2020}{\natexlab{a}})}\BibitemShut {NoStop}%
\bibitem [{\citenamefont {Novario}\ \emph
  {et~al.}(2020{\natexlab{b}})\citenamefont {Novario}, \citenamefont {Gysbers},
  \citenamefont {Engel}, \citenamefont {Hagen}, \citenamefont {Jansen},
  \citenamefont {Morris}, \citenamefont {Navrátil}, \citenamefont
  {Papenbrock},\ and\ \citenamefont {Quaglioni}}]{Novario2020b}%
  \BibitemOpen
  \bibfield  {author} {\bibinfo {author} {\bibfnamefont {S.~J.}\ \bibnamefont
  {Novario}}, \bibinfo {author} {\bibfnamefont {P.}~\bibnamefont {Gysbers}},
  \bibinfo {author} {\bibfnamefont {J.}~\bibnamefont {Engel}}, \bibinfo
  {author} {\bibfnamefont {G.}~\bibnamefont {Hagen}}, \bibinfo {author}
  {\bibfnamefont {G.~R.}\ \bibnamefont {Jansen}}, \bibinfo {author}
  {\bibfnamefont {T.~D.}\ \bibnamefont {Morris}}, \bibinfo {author}
  {\bibfnamefont {P.}~\bibnamefont {Navrátil}}, \bibinfo {author}
  {\bibfnamefont {T.}~\bibnamefont {Papenbrock}}, \ and\ \bibinfo {author}
  {\bibfnamefont {S.}~\bibnamefont {Quaglioni}},\ }\href@noop {} {\enquote
  {\bibinfo {title} {Coupled-cluster calculations of neutrinoless double-beta
  decay in $^{48}$\text{Ca}},}\ } (\bibinfo {year} {2020}{\natexlab{b}}),\
  \Eprint {http://arxiv.org/abs/2008.09696} {arXiv:2008.09696} \BibitemShut
  {NoStop}%
\bibitem [{\citenamefont {Hay}(1973)}]{Hay73natPT}%
  \BibitemOpen
  \bibfield  {author} {\bibinfo {author} {\bibfnamefont {P.~J.}\ \bibnamefont
  {Hay}},\ }\bibfield  {title} {\enquote {\bibinfo {title} {On the calculation
  of natural orbitals by perturbation theory},}\ }\href {\doibase
  10.1063/1.1680359} {\bibfield  {journal} {\bibinfo  {journal} {J. Chem.
  Phys.}\ }\textbf {\bibinfo {volume} {59}},\ \bibinfo {pages} {2468} (\bibinfo
  {year} {1973})}\BibitemShut {NoStop}%
\bibitem [{\citenamefont {Siu}\ and\ \citenamefont
  {Hayes}(1974)}]{Siu74CInatPT}%
  \BibitemOpen
  \bibfield  {author} {\bibinfo {author} {\bibfnamefont {A.~K.~Q.}\
  \bibnamefont {Siu}}\ and\ \bibinfo {author} {\bibfnamefont {E.~F.}\
  \bibnamefont {Hayes}},\ }\bibfield  {title} {\enquote {\bibinfo {title}
  {{Configuration interaction procedure based on the calculation of
  perturbation theory natural orbitals: Applications to {H}$_2$ and {LiH}}},}\
  }\href {\doibase 10.1063/1.1681646} {\bibfield  {journal} {\bibinfo
  {journal} {J. Chem. Phys.}\ }\textbf {\bibinfo {volume} {61}},\ \bibinfo
  {pages} {37} (\bibinfo {year} {1974})}\BibitemShut {NoStop}%
\bibitem [{\citenamefont {Strayer}\ \emph {et~al.}(1973)\citenamefont
  {Strayer}, \citenamefont {Bassichis},\ and\ \citenamefont
  {Kerman}}]{Stra73nucldens}%
  \BibitemOpen
  \bibfield  {author} {\bibinfo {author} {\bibfnamefont {M.~R.}\ \bibnamefont
  {Strayer}}, \bibinfo {author} {\bibfnamefont {W.~H.}\ \bibnamefont
  {Bassichis}}, \ and\ \bibinfo {author} {\bibfnamefont {A.~K.}\ \bibnamefont
  {Kerman}},\ }\bibfield  {title} {\enquote {\bibinfo {title} {{Correlation
  Effects in Nuclear Densities}},}\ }\href {\doibase 10.1103/PhysRevC.8.1269}
  {\bibfield  {journal} {\bibinfo  {journal} {Phys. Rev. C}\ }\textbf {\bibinfo
  {volume} {8}},\ \bibinfo {pages} {1269} (\bibinfo {year} {1973})}\BibitemShut
  {NoStop}%
\bibitem [{\citenamefont {Szabo}\ and\ \citenamefont
  {Ostlund}(1989)}]{Szab89QChem}%
  \BibitemOpen
  \bibfield  {author} {\bibinfo {author} {\bibfnamefont {A.}~\bibnamefont
  {Szabo}}\ and\ \bibinfo {author} {\bibfnamefont {N.S.}\ \bibnamefont
  {Ostlund}},\ }\href@noop {} {\emph {\bibinfo {title} {Modern Quantum
  Chemistry: Introduction to Advanced Electronic Structure Theory}}},\ Dover
  Books on Chemistry\ (\bibinfo  {publisher} {Dover Publications},\ \bibinfo
  {year} {1989})\BibitemShut {NoStop}%
\bibitem [{\citenamefont {Shavitt}\ and\ \citenamefont
  {Bartlett}(2009)}]{Shav09MBmethod}%
  \BibitemOpen
  \bibfield  {author} {\bibinfo {author} {\bibfnamefont {Isaiah}\ \bibnamefont
  {Shavitt}}\ and\ \bibinfo {author} {\bibfnamefont {Rodney~J.}\ \bibnamefont
  {Bartlett}},\ }\href@noop {} {\emph {\bibinfo {title} {Many-Body Methods in
  Chemistry and Physics: MBPT and Coupled-Cluster Theory}}},\ Cambridge
  Molecular Science\ (\bibinfo  {publisher} {Cambridge University Press},\
  \bibinfo {year} {2009})\BibitemShut {NoStop}%
\bibitem [{\citenamefont {Carbone}\ \emph {et~al.}(2013)\citenamefont
  {Carbone}, \citenamefont {Cipollone}, \citenamefont {Barbieri}, \citenamefont
  {Rios},\ and\ \citenamefont {Polls}}]{Carb13SCGF3B}%
  \BibitemOpen
  \bibfield  {author} {\bibinfo {author} {\bibfnamefont {A.}~\bibnamefont
  {Carbone}}, \bibinfo {author} {\bibfnamefont {A.}~\bibnamefont {Cipollone}},
  \bibinfo {author} {\bibfnamefont {C.}~\bibnamefont {Barbieri}}, \bibinfo
  {author} {\bibfnamefont {A.}~\bibnamefont {Rios}}, \ and\ \bibinfo {author}
  {\bibfnamefont {A.}~\bibnamefont {Polls}},\ }\bibfield  {title} {\enquote
  {\bibinfo {title} {{Self-consistent Green's functions formalism with
  three-body interactions}},}\ }\href {\doibase 10.1103/PhysRevC.88.054326}
  {\bibfield  {journal} {\bibinfo  {journal} {Phys. Rev. C}\ }\textbf {\bibinfo
  {volume} {88}},\ \bibinfo {pages} {054326} (\bibinfo {year}
  {2013})}\BibitemShut {NoStop}%
\bibitem [{\citenamefont {Constantinou}\ \emph {et~al.}(2017)\citenamefont
  {Constantinou}, \citenamefont {Caprio}, \citenamefont {Vary},\ and\
  \citenamefont {Maris}}]{Const17NatHe6}%
  \BibitemOpen
  \bibfield  {author} {\bibinfo {author} {\bibfnamefont {C.}~\bibnamefont
  {Constantinou}}, \bibinfo {author} {\bibfnamefont {M.~A.}\ \bibnamefont
  {Caprio}}, \bibinfo {author} {\bibfnamefont {J.~P.}\ \bibnamefont {Vary}}, \
  and\ \bibinfo {author} {\bibfnamefont {P.}~\bibnamefont {Maris}},\ }\bibfield
   {title} {\enquote {\bibinfo {title} {{Natural orbital description of the
  halo nucleus $^{6}$He}},}\ }\href {\doibase 10.1007/s41365-017-0332-6}
  {\bibfield  {journal} {\bibinfo  {journal} {Nucl. Sci. Tech.}\ }\textbf
  {\bibinfo {volume} {28}},\ \bibinfo {pages} {179} (\bibinfo {year}
  {2017})}\BibitemShut {NoStop}%
\bibitem [{\citenamefont {Khadkikar}\ and\ \citenamefont
  {Kamble}(1974)}]{Khad74TkinHF}%
  \BibitemOpen
  \bibfield  {author} {\bibinfo {author} {\bibfnamefont {S.~B.}\ \bibnamefont
  {Khadkikar}}\ and\ \bibinfo {author} {\bibfnamefont {V.~B.}\ \bibnamefont
  {Kamble}},\ }\bibfield  {title} {\enquote {\bibinfo {title} {{Centre of mass
  motion and single particle separation energies in Hartree-Fock theory}},}\
  }\href {\doibase https://doi.org/10.1016/0375-9474(74)90545-4} {\bibfield
  {journal} {\bibinfo  {journal} {Nucl. Phys. A}\ }\textbf {\bibinfo {volume}
  {225}},\ \bibinfo {pages} {352} (\bibinfo {year} {1974})}\BibitemShut
  {NoStop}%
\bibitem [{\citenamefont {Jaqua}\ \emph {et~al.}(1992)\citenamefont {Jaqua},
  \citenamefont {Hasan}, \citenamefont {Vary},\ and\ \citenamefont
  {Barrett}}]{Jaqu92TkinSM}%
  \BibitemOpen
  \bibfield  {author} {\bibinfo {author} {\bibfnamefont {L.}~\bibnamefont
  {Jaqua}}, \bibinfo {author} {\bibfnamefont {M.~A.}\ \bibnamefont {Hasan}},
  \bibinfo {author} {\bibfnamefont {J.~P.}\ \bibnamefont {Vary}}, \ and\
  \bibinfo {author} {\bibfnamefont {B.~R.}\ \bibnamefont {Barrett}},\
  }\bibfield  {title} {\enquote {\bibinfo {title} {Kinetic-energy operator in
  the effective shell-model interaction},}\ }\href {\doibase
  10.1103/PhysRevC.46.2333} {\bibfield  {journal} {\bibinfo  {journal} {Phys.
  Rev. C}\ }\textbf {\bibinfo {volume} {46}},\ \bibinfo {pages} {2333}
  (\bibinfo {year} {1992})}\BibitemShut {NoStop}%
\bibitem [{\citenamefont {Hergert}\ and\ \citenamefont
  {Roth}(2009)}]{Herg09TkinHFB}%
  \BibitemOpen
  \bibfield  {author} {\bibinfo {author} {\bibfnamefont {H.}~\bibnamefont
  {Hergert}}\ and\ \bibinfo {author} {\bibfnamefont {R.}~\bibnamefont {Roth}},\
  }\bibfield  {title} {\enquote {\bibinfo {title} {{Treatment of the intrinsic
  Hamiltonian in particle-number nonconserving theories}},}\ }\href {\doibase
  10.1016/j.physletb.2009.10.100} {\bibfield  {journal} {\bibinfo  {journal}
  {Phys. Lett. B}\ }\textbf {\bibinfo {volume} {682}},\ \bibinfo {pages} {27}
  (\bibinfo {year} {2009})}\BibitemShut {NoStop}%
\bibitem [{\citenamefont {Wick}(1950)}]{Wick50theorem}%
  \BibitemOpen
  \bibfield  {author} {\bibinfo {author} {\bibfnamefont {G.~C.}\ \bibnamefont
  {Wick}},\ }\bibfield  {title} {\enquote {\bibinfo {title} {{The Evaluation of
  the Collision Matrix}},}\ }\href {\doibase 10.1103/PhysRev.80.268} {\bibfield
   {journal} {\bibinfo  {journal} {Phys. Rev.}\ }\textbf {\bibinfo {volume}
  {80}},\ \bibinfo {pages} {268} (\bibinfo {year} {1950})}\BibitemShut
  {NoStop}%
\bibitem [{\citenamefont {Bogner}\ \emph {et~al.}(2010)\citenamefont {Bogner},
  \citenamefont {Furnstahl},\ and\ \citenamefont {Schwenk}}]{Bogn10PPNP}%
  \BibitemOpen
  \bibfield  {author} {\bibinfo {author} {\bibfnamefont {S.~K.}\ \bibnamefont
  {Bogner}}, \bibinfo {author} {\bibfnamefont {R.~J.}\ \bibnamefont
  {Furnstahl}}, \ and\ \bibinfo {author} {\bibfnamefont {A.}~\bibnamefont
  {Schwenk}},\ }\bibfield  {title} {\enquote {\bibinfo {title} {{From
  low-momentum interactions to nuclear structure}},}\ }\href {\doibase
  10.1016/j.ppnp.2010.03.001} {\bibfield  {journal} {\bibinfo  {journal} {Prog.
  Part. Nucl. Phys.}\ }\textbf {\bibinfo {volume} {65}},\ \bibinfo {pages} {94}
  (\bibinfo {year} {2010})}\BibitemShut {NoStop}%
\bibitem [{\citenamefont {Kutzelnigg}\ and\ \citenamefont
  {Mukherjee}(1997)}]{Kutz97mrNO}%
  \BibitemOpen
  \bibfield  {author} {\bibinfo {author} {\bibfnamefont {W.}~\bibnamefont
  {Kutzelnigg}}\ and\ \bibinfo {author} {\bibfnamefont {D.}~\bibnamefont
  {Mukherjee}},\ }\bibfield  {title} {\enquote {\bibinfo {title} {Normal order
  and extended {W}ick theorem for a multiconfiguration reference wave
  function},}\ }\href {\doibase 10.1063/1.474405} {\bibfield  {journal}
  {\bibinfo  {journal} {J. Chem. Phys.}\ }\textbf {\bibinfo {volume} {107}},\
  \bibinfo {pages} {432} (\bibinfo {year} {1997})}\BibitemShut {NoStop}%
\bibitem [{\citenamefont {Gebrerufael}\ \emph {et~al.}(2016)\citenamefont
  {Gebrerufael}, \citenamefont {Calci},\ and\ \citenamefont {Roth}}]{Gebr16MR}%
  \BibitemOpen
  \bibfield  {author} {\bibinfo {author} {\bibfnamefont {E.}~\bibnamefont
  {Gebrerufael}}, \bibinfo {author} {\bibfnamefont {A.}~\bibnamefont {Calci}},
  \ and\ \bibinfo {author} {\bibfnamefont {R.}~\bibnamefont {Roth}},\
  }\bibfield  {title} {\enquote {\bibinfo {title} {{Open-shell nuclei and
  excited states from multireference normal-ordered Hamiltonians}},}\ }\href
  {\doibase 10.1103/PhysRevC.93.031301} {\bibfield  {journal} {\bibinfo
  {journal} {Phys. Rev. C}\ }\textbf {\bibinfo {volume} {93}},\ \bibinfo
  {pages} {031301(R)} (\bibinfo {year} {2016})}\BibitemShut {NoStop}%
\bibitem [{\citenamefont {Hergert}\ \emph {et~al.}(2013)\citenamefont
  {Hergert}, \citenamefont {Binder}, \citenamefont {Calci}, \citenamefont
  {Langhammer},\ and\ \citenamefont {Roth}}]{Herg13MR}%
  \BibitemOpen
  \bibfield  {author} {\bibinfo {author} {\bibfnamefont {H.}~\bibnamefont
  {Hergert}}, \bibinfo {author} {\bibfnamefont {S.}~\bibnamefont {Binder}},
  \bibinfo {author} {\bibfnamefont {A.}~\bibnamefont {Calci}}, \bibinfo
  {author} {\bibfnamefont {J.}~\bibnamefont {Langhammer}}, \ and\ \bibinfo
  {author} {\bibfnamefont {R.}~\bibnamefont {Roth}},\ }\bibfield  {title}
  {\enquote {\bibinfo {title} {{Ab Initio Calculations of Even Oxygen Isotopes
  with Chiral Two-Plus-Three-Nucleon Interactions}},}\ }\href {\doibase
  10.1103/PhysRevLett.110.242501} {\bibfield  {journal} {\bibinfo  {journal}
  {Phys. Rev. Lett.}\ }\textbf {\bibinfo {volume} {110}},\ \bibinfo {pages}
  {242501} (\bibinfo {year} {2013})}\BibitemShut {NoStop}%
\bibitem [{\citenamefont {Gebrerufael}\ \emph {et~al.}(2017)\citenamefont
  {Gebrerufael}, \citenamefont {Vobig}, \citenamefont {Hergert},\ and\
  \citenamefont {Roth}}]{Gebr17IMNCSM}%
  \BibitemOpen
  \bibfield  {author} {\bibinfo {author} {\bibfnamefont {E.}~\bibnamefont
  {Gebrerufael}}, \bibinfo {author} {\bibfnamefont {K.}~\bibnamefont {Vobig}},
  \bibinfo {author} {\bibfnamefont {H.}~\bibnamefont {Hergert}}, \ and\
  \bibinfo {author} {\bibfnamefont {R.}~\bibnamefont {Roth}},\ }\bibfield
  {title} {\enquote {\bibinfo {title} {{Ab Initio Description of Open-Shell
  Nuclei: Merging No-Core Shell Model and In-Medium Similarity Renormalization
  Group}},}\ }\href {\doibase 10.1103/PhysRevLett.118.152503} {\bibfield
  {journal} {\bibinfo  {journal} {Phys. Rev. Lett.}\ }\textbf {\bibinfo
  {volume} {118}},\ \bibinfo {pages} {152503} (\bibinfo {year}
  {2017})}\BibitemShut {NoStop}%
\bibitem [{\citenamefont {Hagen}\ \emph {et~al.}(2007)\citenamefont {Hagen},
  \citenamefont {Papenbrock}, \citenamefont {Dean}, \citenamefont {Schwenk},
  \citenamefont {Nogga}, \citenamefont {W\l{}och},\ and\ \citenamefont
  {Piecuch}}]{Hage07CC3N}%
  \BibitemOpen
  \bibfield  {author} {\bibinfo {author} {\bibfnamefont {G.}~\bibnamefont
  {Hagen}}, \bibinfo {author} {\bibfnamefont {T.}~\bibnamefont {Papenbrock}},
  \bibinfo {author} {\bibfnamefont {D.~J.}\ \bibnamefont {Dean}}, \bibinfo
  {author} {\bibfnamefont {A.}~\bibnamefont {Schwenk}}, \bibinfo {author}
  {\bibfnamefont {A.}~\bibnamefont {Nogga}}, \bibinfo {author} {\bibfnamefont
  {M.}~\bibnamefont {W\l{}och}}, \ and\ \bibinfo {author} {\bibfnamefont
  {P.}~\bibnamefont {Piecuch}},\ }\bibfield  {title} {\enquote {\bibinfo
  {title} {{Coupled-cluster theory for three-body Hamiltonians}},}\ }\href
  {\doibase 10.1103/PhysRevC.76.034302} {\bibfield  {journal} {\bibinfo
  {journal} {Phys. Rev. C}\ }\textbf {\bibinfo {volume} {76}},\ \bibinfo
  {pages} {034302} (\bibinfo {year} {2007})}\BibitemShut {NoStop}%
\bibitem [{\citenamefont {Roth}\ \emph {et~al.}(2012)\citenamefont {Roth},
  \citenamefont {Binder}, \citenamefont {Vobig}, \citenamefont {Calci},
  \citenamefont {Langhammer},\ and\ \citenamefont
  {Navr\'{a}til}}]{Roth12NCSMCC3N}%
  \BibitemOpen
  \bibfield  {author} {\bibinfo {author} {\bibfnamefont {R.}~\bibnamefont
  {Roth}}, \bibinfo {author} {\bibfnamefont {S.}~\bibnamefont {Binder}},
  \bibinfo {author} {\bibfnamefont {K.}~\bibnamefont {Vobig}}, \bibinfo
  {author} {\bibfnamefont {A.}~\bibnamefont {Calci}}, \bibinfo {author}
  {\bibfnamefont {J.}~\bibnamefont {Langhammer}}, \ and\ \bibinfo {author}
  {\bibfnamefont {P.}~\bibnamefont {Navr\'{a}til}},\ }\bibfield  {title}
  {\enquote {\bibinfo {title} {{Medium-Mass Nuclei with Normal-Ordered Chiral
  NN+3N Interactions}},}\ }\href {\doibase 10.1103/PhysRevLett.109.052501}
  {\bibfield  {journal} {\bibinfo  {journal} {Phys. Rev. Lett.}\ }\textbf
  {\bibinfo {volume} {109}},\ \bibinfo {pages} {052501} (\bibinfo {year}
  {2012})}\BibitemShut {NoStop}%
\bibitem [{\citenamefont {Hergert}(2017)}]{Herg17PS}%
  \BibitemOpen
  \bibfield  {author} {\bibinfo {author} {\bibfnamefont {H.}~\bibnamefont
  {Hergert}},\ }\bibfield  {title} {\enquote {\bibinfo {title} {{In-medium
  similarity renormalization group for closed and open-shell nuclei}},}\ }\href
  {\doibase 10.1088/1402-4896/92/2/023002} {\bibfield  {journal} {\bibinfo
  {journal} {Phys. Scr.}\ }\textbf {\bibinfo {volume} {92}},\ \bibinfo {pages}
  {023002} (\bibinfo {year} {2017})}\BibitemShut {NoStop}%
\bibitem [{\citenamefont {Stroberg}\ \emph {et~al.}(2017)\citenamefont
  {Stroberg}, \citenamefont {Calci}, \citenamefont {Hergert}, \citenamefont
  {Holt}, \citenamefont {Bogner}, \citenamefont {Roth},\ and\ \citenamefont
  {Schwenk}}]{Stro17ENO}%
  \BibitemOpen
  \bibfield  {author} {\bibinfo {author} {\bibfnamefont {S.~R.}\ \bibnamefont
  {Stroberg}}, \bibinfo {author} {\bibfnamefont {A.}~\bibnamefont {Calci}},
  \bibinfo {author} {\bibfnamefont {H.}~\bibnamefont {Hergert}}, \bibinfo
  {author} {\bibfnamefont {J.~D.}\ \bibnamefont {Holt}}, \bibinfo {author}
  {\bibfnamefont {S.~K.}\ \bibnamefont {Bogner}}, \bibinfo {author}
  {\bibfnamefont {R.}~\bibnamefont {Roth}}, \ and\ \bibinfo {author}
  {\bibfnamefont {A.}~\bibnamefont {Schwenk}},\ }\bibfield  {title} {\enquote
  {\bibinfo {title} {{Nucleus-Dependent Valence-Space Approach to Nuclear
  Structure}},}\ }\href {\doibase 10.1103/PhysRevLett.118.032502} {\bibfield
  {journal} {\bibinfo  {journal} {Phys. Rev. Lett.}\ }\textbf {\bibinfo
  {volume} {118}},\ \bibinfo {pages} {032502} (\bibinfo {year}
  {2017})}\BibitemShut {NoStop}%
\bibitem [{\citenamefont {Magnus}(1954)}]{Magn54exp}%
  \BibitemOpen
  \bibfield  {author} {\bibinfo {author} {\bibfnamefont {W.}~\bibnamefont
  {Magnus}},\ }\bibfield  {title} {\enquote {\bibinfo {title} {{On the
  exponential solution of differential equations for a linear operator}},}\
  }\href {\doibase 10.1002/cpa.3160070404} {\bibfield  {journal} {\bibinfo
  {journal} {Commun. Pure Appl. Math.}\ }\textbf {\bibinfo {volume} {7}},\
  \bibinfo {pages} {649} (\bibinfo {year} {1954})}\BibitemShut {NoStop}%
\bibitem [{\citenamefont {Morris}\ \emph {et~al.}(2015)\citenamefont {Morris},
  \citenamefont {Parzuchowski},\ and\ \citenamefont {Bogner}}]{Morr15Magnus}%
  \BibitemOpen
  \bibfield  {author} {\bibinfo {author} {\bibfnamefont {T.~D.}\ \bibnamefont
  {Morris}}, \bibinfo {author} {\bibfnamefont {N.~M.}\ \bibnamefont
  {Parzuchowski}}, \ and\ \bibinfo {author} {\bibfnamefont {S.~K.}\
  \bibnamefont {Bogner}},\ }\bibfield  {title} {\enquote {\bibinfo {title}
  {{Magnus expansion and in-medium similarity renormalization group}},}\ }\href
  {\doibase 10.1103/PhysRevC.92.034331} {\bibfield  {journal} {\bibinfo
  {journal} {Phys. Rev. C}\ }\textbf {\bibinfo {volume} {92}},\ \bibinfo
  {pages} {034331} (\bibinfo {year} {2015})}\BibitemShut {NoStop}%
\bibitem [{\citenamefont {Furnstahl}(2012)}]{Furn12NPPS}%
  \BibitemOpen
  \bibfield  {author} {\bibinfo {author} {\bibfnamefont {R.~J.}\ \bibnamefont
  {Furnstahl}},\ }\bibfield  {title} {\enquote {\bibinfo {title} {{The
  Renormalization Group in Nuclear Physics}},}\ }\href {\doibase
  10.1016/j.nuclphysbps.2012.06.005} {\bibfield  {journal} {\bibinfo  {journal}
  {Nucl. Phys. Proc. Suppl.}\ }\textbf {\bibinfo {volume} {228}},\ \bibinfo
  {pages} {139} (\bibinfo {year} {2012})}\BibitemShut {NoStop}%
\bibitem [{\citenamefont {Bogner}\ \emph {et~al.}(2006)\citenamefont {Bogner},
  \citenamefont {Furnstahl}, \citenamefont {Ramanan},\ and\ \citenamefont
  {Schwenk}}]{Bogn06bseries}%
  \BibitemOpen
  \bibfield  {author} {\bibinfo {author} {\bibfnamefont {S.~K.}\ \bibnamefont
  {Bogner}}, \bibinfo {author} {\bibfnamefont {R.~J.}\ \bibnamefont
  {Furnstahl}}, \bibinfo {author} {\bibfnamefont {S.}~\bibnamefont {Ramanan}},
  \ and\ \bibinfo {author} {\bibfnamefont {A.}~\bibnamefont {Schwenk}},\
  }\bibfield  {title} {\enquote {\bibinfo {title} {{Convergence of the Born
  series with low-momentum interactions}},}\ }\href {\doibase
  10.1016/j.nuclphysa.2006.05.004} {\bibfield  {journal} {\bibinfo  {journal}
  {Nucl. Phys. A}\ }\textbf {\bibinfo {volume} {773}},\ \bibinfo {pages} {203}
  (\bibinfo {year} {2006})}\BibitemShut {NoStop}%
\bibitem [{\citenamefont {Hoppe}\ \emph {et~al.}(2017)\citenamefont {Hoppe},
  \citenamefont {Drischler}, \citenamefont {Furnstahl}, \citenamefont
  {Hebeler},\ and\ \citenamefont {Schwenk}}]{Hopp17WeinEVAn}%
  \BibitemOpen
  \bibfield  {author} {\bibinfo {author} {\bibfnamefont {J.}~\bibnamefont
  {Hoppe}}, \bibinfo {author} {\bibfnamefont {C.}~\bibnamefont {Drischler}},
  \bibinfo {author} {\bibfnamefont {R.~J.}\ \bibnamefont {Furnstahl}}, \bibinfo
  {author} {\bibfnamefont {K.}~\bibnamefont {Hebeler}}, \ and\ \bibinfo
  {author} {\bibfnamefont {A.}~\bibnamefont {Schwenk}},\ }\bibfield  {title}
  {\enquote {\bibinfo {title} {Weinberg eigenvalues for chiral nucleon-nucleon
  interactions},}\ }\href {\doibase 10.1103/PhysRevC.96.054002} {\bibfield
  {journal} {\bibinfo  {journal} {Phys. Rev. C}\ }\textbf {\bibinfo {volume}
  {96}},\ \bibinfo {pages} {054002} (\bibinfo {year} {2017})}\BibitemShut
  {NoStop}%
\bibitem [{\citenamefont {Simonis}\ \emph {et~al.}(2017)\citenamefont
  {Simonis}, \citenamefont {Stroberg}, \citenamefont {Hebeler}, \citenamefont
  {Holt},\ and\ \citenamefont {Schwenk}}]{Simo17SatFinNuc}%
  \BibitemOpen
  \bibfield  {author} {\bibinfo {author} {\bibfnamefont {J.}~\bibnamefont
  {Simonis}}, \bibinfo {author} {\bibfnamefont {S.~R.}\ \bibnamefont
  {Stroberg}}, \bibinfo {author} {\bibfnamefont {K.}~\bibnamefont {Hebeler}},
  \bibinfo {author} {\bibfnamefont {J.~D.}\ \bibnamefont {Holt}}, \ and\
  \bibinfo {author} {\bibfnamefont {A.}~\bibnamefont {Schwenk}},\ }\bibfield
  {title} {\enquote {\bibinfo {title} {{Saturation with chiral interactions and
  consequences for finite nuclei}},}\ }\href {\doibase
  10.1103/PhysRevC.96.014303} {\bibfield  {journal} {\bibinfo  {journal} {Phys.
  Rev. C}\ }\textbf {\bibinfo {volume} {96}},\ \bibinfo {pages} {014303}
  (\bibinfo {year} {2017})}\BibitemShut {NoStop}%
\bibitem [{\citenamefont {Angeli}\ and\ \citenamefont
  {Marinova}(2013)}]{Ange13rch}%
  \BibitemOpen
  \bibfield  {author} {\bibinfo {author} {\bibfnamefont {I.}~\bibnamefont
  {Angeli}}\ and\ \bibinfo {author} {\bibfnamefont {K.~P.}\ \bibnamefont
  {Marinova}},\ }\bibfield  {title} {\enquote {\bibinfo {title} {Table of
  experimental nuclear ground state charge radii: An update},}\ }\href
  {\doibase 10.1016/j.adt.2011.12.006} {\bibfield  {journal} {\bibinfo
  {journal} {At. Data Nucl. Data Tables}\ }\textbf {\bibinfo {volume} {99}},\
  \bibinfo {pages} {69} (\bibinfo {year} {2013})}\BibitemShut {NoStop}%
\bibitem [{\citenamefont {Wang}\ \emph {et~al.}(2017)\citenamefont {Wang},
  \citenamefont {Audi}, \citenamefont {Kondev}, \citenamefont {Huang},
  \citenamefont {Naimi},\ and\ \citenamefont {Xu}}]{Wang17AME16}%
  \BibitemOpen
  \bibfield  {author} {\bibinfo {author} {\bibfnamefont {M.}~\bibnamefont
  {Wang}}, \bibinfo {author} {\bibfnamefont {G.}~\bibnamefont {Audi}}, \bibinfo
  {author} {\bibfnamefont {F.~G.}\ \bibnamefont {Kondev}}, \bibinfo {author}
  {\bibfnamefont {W.~J.}\ \bibnamefont {Huang}}, \bibinfo {author}
  {\bibfnamefont {S.}~\bibnamefont {Naimi}}, \ and\ \bibinfo {author}
  {\bibfnamefont {X.}~\bibnamefont {Xu}},\ }\bibfield  {title} {\enquote
  {\bibinfo {title} {The {AME}2016 atomic mass evaluation ({II}). {T}ables,
  graphs and references},}\ }\href {\doibase 10.1088/1674-1137/41/3/030003}
  {\bibfield  {journal} {\bibinfo  {journal} {Chin. Phys. C}\ }\textbf
  {\bibinfo {volume} {41}},\ \bibinfo {pages} {030003} (\bibinfo {year}
  {2017})}\BibitemShut {NoStop}%
\bibitem [{\citenamefont {Gordon}\ \emph {et~al.}(1999)\citenamefont {Gordon},
  \citenamefont {Schmidt}, \citenamefont {Chaban}, \citenamefont {Glaesemann},
  \citenamefont {Stevens},\ and\ \citenamefont {Gonzalez}}]{Gord99negNAT}%
  \BibitemOpen
  \bibfield  {author} {\bibinfo {author} {\bibfnamefont {M.~S.}\ \bibnamefont
  {Gordon}}, \bibinfo {author} {\bibfnamefont {M.~W.}\ \bibnamefont {Schmidt}},
  \bibinfo {author} {\bibfnamefont {G.~M.}\ \bibnamefont {Chaban}}, \bibinfo
  {author} {\bibfnamefont {K.~R.}\ \bibnamefont {Glaesemann}}, \bibinfo
  {author} {\bibfnamefont {W.~J.}\ \bibnamefont {Stevens}}, \ and\ \bibinfo
  {author} {\bibfnamefont {C.}~\bibnamefont {Gonzalez}},\ }\bibfield  {title}
  {\enquote {\bibinfo {title} {A natural orbital diagnostic for
  multiconfigurational character in correlated wave functions},}\ }\href
  {\doibase 10.1063/1.478301} {\bibfield  {journal} {\bibinfo  {journal} {J.
  Chem. Phys.}\ }\textbf {\bibinfo {volume} {110}},\ \bibinfo {pages} {4199}
  (\bibinfo {year} {1999})}\BibitemShut {NoStop}%
\bibitem [{\citenamefont {Hebeler}(2012)}]{Hebe12msSRG}%
  \BibitemOpen
  \bibfield  {author} {\bibinfo {author} {\bibfnamefont {K.}~\bibnamefont
  {Hebeler}},\ }\bibfield  {title} {\enquote {\bibinfo {title} {{Momentum-space
  evolution of chiral three-nucleon forces}},}\ }\href {\doibase
  10.1103/PhysRevC.85.021002} {\bibfield  {journal} {\bibinfo  {journal} {Phys.
  Rev. C}\ }\textbf {\bibinfo {volume} {85}},\ \bibinfo {pages} {021002(R)}
  (\bibinfo {year} {2012})}\BibitemShut {NoStop}%
\bibitem [{\citenamefont {Stroberg}(Nov 23, 2019)}]{Stro17imsrggit}%
  \BibitemOpen
  \bibfield  {author} {\bibinfo {author} {\bibfnamefont {S.~R.}\ \bibnamefont
  {Stroberg}},\ }\href@noop {} {\bibfield  {journal} {\bibinfo  {journal}
  {{https://github.com/ragnarstroberg/imsrg}}\ } (\bibinfo {year} {Nov 23,
  2019})}\BibitemShut {NoStop}%
\end{thebibliography}%

\end{document}